\documentclass[twocolumn,superscriptaddress,floatfix,preprintnumbers, nofootinbib,hyperref]{revtex4-2} 
\pdfoutput=1
\usepackage[colorlinks=true,breaklinks=true]{hyperref}
\usepackage[normalem]{ulem}
\usepackage[utf8]{inputenc}
\hypersetup{allcolors=[rgb]{0.0 0.0 0.6},linkcolor=[rgb]{0.75 0.05 0.05}}
\usepackage{amsmath,amssymb}
\usepackage{epsfig}  
\usepackage{graphicx}   
\usepackage{slashed}       
\usepackage{tikz}
\usepackage{url}
\usepackage{color}
\usepackage{multirow}
\usepackage{orcidlink}
\usepackage{comment}
\usepackage{amssymb}

\usepackage{empheq,etoolbox}
\patchcmd{\subequations}
  {\theparentequation\alph{equation}}
  {\theparentequation.\alph{equation}}
  {}{}
  
\DeclareMathOperator{\cm}{cm}
\DeclareMathOperator{\GeV}{GeV}
\DeclareMathOperator{\eV}{eV}

\DeclareMathOperator{\MeV}{MeV}

\DeclareMathOperator{\Mpc}{Mpc}

\newcommand{\beq}{\begin{equation}}
\newcommand{\eeq}{\end{equation}}

\allowdisplaybreaks

\bibpunct{[}{]}{,}{n}{}{,}

\begin{document}

\title{Probing the Blue Axion with Cosmic Optical Background Anisotropies
}

\author{Pierluca Carenza \orcidlink{0000-0002-8410-0345}}\email{pierluca.carenza@fysik.su.se}
\affiliation{The Oskar Klein Centre, Department of Physics, Stockholm University, 10691 Stockholm, Sweden}

\author{Giuseppe Lucente \orcidlink{0000-0003-1530-4851}}
\email{giuseppe.lucente@ba.infn.it}
\affiliation{Dipartimento Interateneo di Fisica “Michelangelo Merlin”, Via Amendola 173, 70126 Bari, Italy}
\affiliation{Istituto Nazionale di Fisica Nucleare -- Sezione di Bari, Via Orabona 4, 70126 Bari, Italy}%

\author{Edoardo Vitagliano \orcidlink{0000-0001-7847-1281}}
\email{edoardo.vitagliano@mail.huji.ac.il}
\affiliation{Racah Institute of Physics, Hebrew University of Jerusalem, Jerusalem 91904, Israel}

\smallskip

\begin{abstract}
A radiative decaying Big Bang relic with a mass $m_a\simeq 5-25 \,\rm eV$, which we dub ``blue axion'', can be probed with direct and indirect observations of the cosmic optical background (COB). The strongest bounds on blue-axion cold dark matter come from the Hubble Space Telescope (HST) measurements of COB anisotropies at $606$~nm. We suggest that new HST measurements at higher frequencies ($336$~nm and $438$~nm) can improve current constraints on the lifetime up to one order of magnitude, and we show that also thermally produced and hot relic blue axions can be competitively probed by COB anisotropies. We exclude the simple interpretation of the excess in the diffuse COB detected by the Long Range Reconnaissance Imager (LORRI) as photons produced by a decaying hot relic. Finally, we comment on the reach of upcoming line intensity mapping experiments, that could detect blue axions with a lifetime as large as $10^{29}\,\rm s$ or $10^{27}\,\rm s$ for the cold dark matter and the hot relic case, respectively.

\end{abstract}

\maketitle

\section{Introduction}
The extragalactic background light (EBL) is the accumulated radiation emitted in the Universe by all galaxies and active galactic nuclei over the cosmic history, ranging from far-infrared to ultraviolet bands~\cite{Ressell:1989rz,Dwek:2012nb,Cooray:2016jrk,Hill:2018trh,Biteau:2022dtt}. 
A robust lower bound on the cosmic optical background (COB) is obtained through the Hubble Space Telescope (HST) galaxy counts~\cite{Driver:2016krv,Saldana-Lopez:2020qzx}. However, the shape and intensity of the COB spectrum contributions from diffuse, unresolved sources need to be obtained with different strategies---e.g. by directly detecting the COB with a spacecraft, so to avoid
the airglow and artificial light affecting ground-based
measurements. The Long Range Reconnaissance Imager (LORRI) instrument on NASA's New Horizons mission  has reported the COB to be $16.37\pm1.47\, {\rm nW} \, {\rm m}^{-2}\,  {\rm sr}^{-1}$ at the pivot wavelength of $0.608\rm\, \mu m$~\cite{Zemcov:2017dwy,Lauer:2020qwk,Lauer:2022fgc}. This is about $\sim 4\sigma$ above the HST galaxy count estimate,
suggesting the existence of an unaccounted for EBL component of $8.06 \pm 1.92\, \rm nW\,m^{-2}\, sr^{-1}$. Such excess could be due to e.g. high redshift galaxies~\cite{Yue:2012dd} or direct-collapse black holes~\cite{Yue:2013hya}. A more mundane explanation of the excess could be related to the modeling of the zodiacal light (ZL), sunlight scattered by interplanetary dust. Zodiacal light largely dominates the sky brightness in the inner solar system. While it can be neglected at the distances from the Sun (51.3~A.U.) where LORRI observations
were obtained, a careful estimate of ZL is necessary to evaluate the contribution of diffuse galactic light to the total sky~\cite{Lauer:2022fgc}.

Since $\gamma$-rays (around $100\, \rm GeV$) are absorbed through electron-positron pair
production when scattering on $\mathcal{O}(10\,\rm eV)$ COB photons, analyses of the observed blazar spectra allow for an indirect measurement of the COB and its redshift evolution~\cite{Dwek:2012nb,Cooray:2016jrk,Hill:2018trh,Biteau:2022dtt}. This approach has other systematic uncertainties that hinder an easy evaluation of the COB, most notably the knowledge of the injected blazar spectrum, and the possible production of secondary $\gamma$-rays~\cite{Dwek:2012nb,Essey:2009ju,Essey:2010er,Aharonian:2012fu}. Thus, several effects can potentially undermine the indirect estimates, which are found to be comparable to the COB inferred from galaxy counts~\cite{Fermi-LAT:2018lqt,Desai:2019pfl}, and in tension with direct measurements. (See however Ref.~\cite{MAGIC:2019ozu} for an estimate with larger uncertainties.)

An alternative to the above mentioned approaches relies on measuring the anisotropies rather than the diffuse intensity of the COB. The foregrounds such
as ZL have smooth spatial distributions
and a different correlation function compared to the fluctuations generated by hypothetical extragalactic signals, therefore tackling one of the major drawbacks of direct measurements~\cite{Zemcov:2014eca,Mitchell-Wynne:2015rha,Gong:2015hke}. 

\begin{figure*}
	\includegraphics[width=0.8\textwidth]{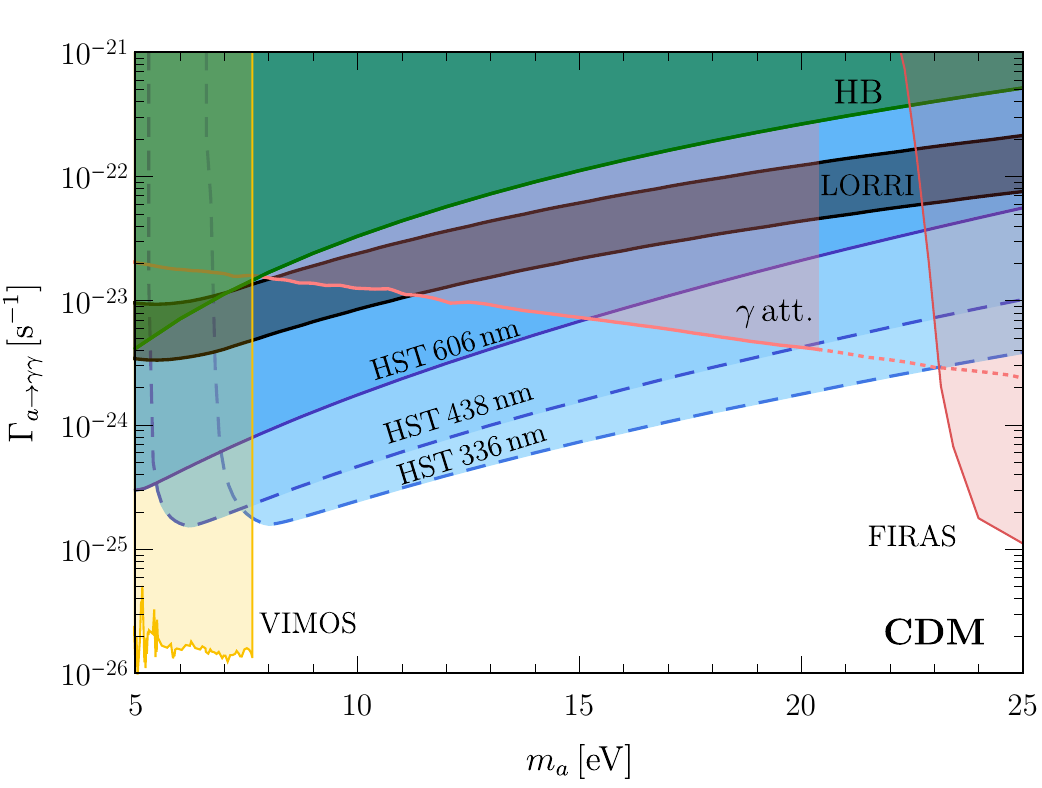}
	\caption{Bounds and projected reaches on the lifetime of the blue-axion cold dark matter. We recompute the $606\,\rm nm$ bound (solid blue lines), and find good agreement with the results of Ref.~\cite{Nakayama:2022jza}. The forecasts of our proposed observations at $438\,\rm nm$ and $336\,\rm nm$ are given with dashed blue lines. Other constraints are also shown: HB stars~\cite{Ayala:2014pea} (green), VIMOS~\cite{Grin:2006aw} (yellow), FIRAS~\cite{Bolliet:2020ofj} (red), and $\gamma$-ray attenuation~\cite{Bernal:2022xyi} (pink) bounds (see Appendix~\ref{app:bounds}). 
	The black band identifies the $95\%$ CL excess detected by LORRI~\cite{Lauer:2022fgc,Bernal:2022wsu}.} 
	\label{fig:boundratecdm}
\end{figure*}

Big Bang relics $a$ decaying to photons through processes such as $a \rightarrow \gamma+\gamma$ and $a\rightarrow \chi+\gamma$ can potentially contribute to the EBL, and crucially to the COB if they have a mass $m_a\simeq 5-25 \,\rm eV$. The authors of Ref.~\cite{Bernal:2022wsu} have shown that the excess in the diffuse COB detected by LORRI can be produced by dark matter (DM) in the form of axion-like particles decaying to blue and ultraviolet light, hereafter dubbed ``blue axions''. However, HST anisotropy measurements at $606$~nm already exclude the hypothesis that the excess is due to cold blue axions~\cite{Nakayama:2022jza}. In the following, we update the cold dark matter (CDM) anisotropy bound from $606$~nm measurements with an improved analysis, which includes two additional anisotropy sources (shot noise and foregrounds) and a refined treatment of the detector response, finding good agreement with Ref.~\cite{Nakayama:2022jza}. We will show that new dedicated HST measurements at wavelengths $336 \,\rm nm$ and $438 \,\rm nm$ can constrain the lifetime of the blue axion by a further factor 4 to 10, resulting in the best probe to date, as can be seen in Fig.~\ref{fig:boundratecdm}. Additional shorter
wavelength anisotropy measurements will be targets
for forthcoming sub-orbital and space-based measurements~\cite{Symons:2022lke}.

The angular power spectrum depends on the abundance of the relic $\rho_a$ and its power spectrum $P_\delta$, therefore the same formalism can be applied to different production mechanism scenarios once $\rho_a$ and $P_\delta$ are evaluated for each case. Besides the cold blue axions produced through misalignment mechanism, we will show that $336 \,\rm nm$ and $438 \,\rm nm$ anisotropy measurements can competitively probe blue axions produced from the annihilation of string-wall networks arising from a symmetry breaking pattern. In this case, they can constitute 100\% of the DM but could have a cutoff in the power spectrum, similar to warm dark matter. Blue axions could also be a non-cold dark matter (NCDM) component. Such a component could have either the temperature equal to neutrino temperature (``hot relic'') or a temperature set by a freeze-out mechanism in cosmological scenarios. In the latter case,  additional degrees of freedom above the Electroweak scale are required to satisfy structure formation bounds on NCDM abundance~\cite{Diamanti:2017xfo}. Nevertheless, we will show that even a hotter component is excluded by anisotropy measurements, further constraining the interpretation of LORRI excess as a decaying Big Bang relic. 

We also tackle the problem of alternative particle-physics inspired interpretations of an excess in the diffuse COB. Any such interpretation needs to avoid anisotropy constraints. Moreover, if blue axions decay to two photons ($a\rightarrow \gamma+\gamma$), strong bounds arise from globular cluster observations, since they would be copiously produced in horizontal branch (HB) stars through Primakoff production~\cite{Ayala:2014pea}. An intuitive way to circumvent both anisotropy and stellar cooling bounds is granted by a small hot dark matter component decaying through a dark portal featuring an additional dark vector, $a\rightarrow \chi+\gamma$~\cite{Kalashev:2018bra}.
In this case, the power spectrum is suppressed at small scales, and the stellar cooling bound is given by the required agreement between the predicted and
observationally inferred core mass at the helium flash of red giants (RGs), since dark sector particles can be produced by plasmon decay $\gamma^*\rightarrow \chi+a$. 

Finally, we explore the possibilities offered by line intensity mapping (LIM),  an emerging tool for cosmology with the goal of measuring the integrated emission along the line of sight from spectral lines emitted in the past. A decaying relic would show up as an ``interloper line'', and applying simple scaling arguments to recent results in the literature we show that LIM can probe blue axion hot relics with a lifetime as large as $10^{27}\,\rm s$.

We begin in Sec.~\ref{sec:bcgk} with a derivation of the general anisotropy power spectrum due to a decaying Big Bang Relic. Then in Sec.~\ref{sec:production} we present the possible production mechanisms of the blue axion, and the corresponding bounds and reaches from anisotropy measurements are discussed in Sec.~\ref{sec:results}. In Sec.~\ref{sec:dark} we explore the dark portal scenario. Sec.~\ref{sec:LIM} is dedicated to the projected reach of LIM experiments. Finally, Sec.~\ref{sec:conclusions} is devoted to a summary and discussion.

\section{Isotropic and anisotropic cosmic optical background}\label{sec:bcgk}

We assume that a population of blue axions exists, and for the time being we will be agnostic about its production mechanism and abundance. Blue axions are assumed to have a coupling to two photons,
 \begin{align}\label{eq:lag}
     \mathcal{L}\supset \frac{1}{4}g_{a\gamma \gamma} a F_{\mu\nu}\tilde{F}^{\mu\nu},
 \end{align}
 where $F_{\mu\nu}=\partial_\mu A_\nu-\partial_\nu A_\mu$ and $\tilde{F}_{\mu\nu}=\epsilon_{\rho\sigma\mu\nu}F^{\rho\sigma}/2$.
 Because of this interaction Lagrangian, blue axions  decay to two photons with a decay rate
\begin{equation}
   \Gamma_{a\to\gamma\gamma}= \frac{ g_{a\gamma\gamma}^2}{64\pi} m_{a}^{3}.
\end{equation}
Following Ref. \cite{Kalashev:2018bra} to compute the contribution to the COB from the decaying blue axions,
we define the average energy intensity in units of energy per time per surface per steradians
\begin{equation}
    \langle I(\omega)\rangle=\frac{\omega^{2}}{4\pi}\frac{dN_{\gamma}}{dS\,d\omega\,dt}=\omega^2\int dz\, W\left[\omega(1+z),z\right],
\label{eq:window}
\end{equation}
where we also introduced the window function $W$.

If one assumes that the relic decays at rest, the decay spectrum of photons is monochromatic with energy $\omega_{\rm max}=m_a/2$ and the energy gets redshifted. The intensity is obtained integrating over the cosmic time and accounting for $\zeta=2$ photons produced in each decay~\cite{Cirelli:2009dv,Cirelli:2010xx,Kalashev:2018bra},
\begin{align}
        \langle I(\omega)\rangle&=\frac{\omega^{2}}{4\pi}\int_0^\infty \frac{dz}{H(z)}\,\frac{\rho_a}{m_a} \zeta\Gamma_{a\to\gamma\gamma}\delta\left[\omega(1+z)-\omega_{\rm max} \right]
        \nonumber
        \\
        &=
    \frac{\omega}{4\pi}\frac{\rho_a}{m_a}\frac{\zeta\Gamma_{a\to\gamma\gamma}}{H (\tilde{z})},
\label{eq:aveint}
\end{align}

where $\tilde{z}=\frac{\omega_{\rm max}}{\omega}-1$. The Hubble parameter at the redshift $z$ is given by $H(z)=H_0 \sqrt{\Omega_{\Lambda}+ \Omega_{m} (1+z)^3 +\Omega_{r} (1+z)^4}$ where $H_0=h\, 100\rm \, km\, s^{-1}\, Mpc^{-1}$ is the value of the Hubble parameter today, $h=0.674$, and $\Omega_{\Lambda}=0.685$, $\Omega_{m}=0.315$ and $\Omega_{r}=5.38\times 10^{-5}$ (neglecting the contribution of massless neutrinos) denote the density parameter of the dark energy, total matter and radiation, respectively, taken from Ref.~\cite{ParticleDataGroup:2022pth}. Comparing Eq.~\eqref{eq:window} to Eq.~\eqref{eq:aveint}, the explicit form of the window function $W$ is obtained. We assume that the depletion in the blue axion population due to decay is negligible, and that the decay happens when blue axions are non-relativistic. We see that, to explain the excess detected by LORRI, assuming blue axions to constitute 100\% of the dark matter and $\omega\simeq m_a$, the rate needs to be $\Gamma_{a\to \gamma\gamma}\simeq 10^{-23}-10^{-22}\, \rm s^{-1}$.

Since blue axions can be clumped, their decay can show up anisotropically in the sky.
To discuss the anisotropy, we first need to define the energy intensity as seen in the detector while pointing at the direction in the sky $\hat{\mathbf{n}}$,
\begin{align}
    I(\omega_{\rm piv}^2, \hat{\mathbf{n}})=\omega_{\rm piv}^2\int \frac{d\omega}{\omega} \int dz\, W\left[\omega(1+z),z,\hat{\mathbf{n}}\right] \epsilon(\omega),
\end{align}
where $\omega_{\rm piv}$ is the pivot frequency. We define $\epsilon(\omega)$ as a normalized throughput function, as we will compare the blue axion anisotropy spectrum to the true anisotropy spectrum, rather than the spectrum as seen in the detector, the Wide Field Camera 3 on board of the Hubble Space Telescope~\cite{WFC3}.\footnote{Further details on HST can be found in Appendix~\ref{app:HST}.} Therefore, calling $\mathbb{T}(\omega)$ the throughput of Ref.~\cite{HST}, which is defined as the number of detected counts/s/$\rm cm^2$ of telescope area relative to the incident flux in photons/s/$\rm cm^2$, our normalized throughput function will be
\begin{align}
    \epsilon(\omega)=\frac{\mathbb{T}(\omega)}{\int_0^\infty \frac{d\omega}{\omega} \mathbb{T}(\omega)},
\end{align}
where the integral in the denominator is the efficiency as defined in Ref.~\cite{WFC3}.
 The fluctuations towards $\hat{\mathbf{n}}$ can
be expanded as spherical harmonics,
\begin{align}
	\delta I (\omega_{\rm piv},\hat{\mathbf{n}})  &= I (\omega_{\rm piv},\hat{\mathbf{n}}) - \langle{I}(\omega_{\rm piv})\rangle \nonumber \\
	&= \sum_{l, m} a_{l m}(\omega_{\rm piv})Y_{l m}(\hat{\mathbf{n}}),
\end{align}
while the relevant angular power spectrum is defined as \begin{equation}
C_l(\omega_{\rm piv})=\langle|a_{l m}(\omega_{\rm piv})|^2\rangle=\frac{1}{2l+1}\sum_{m=-l}^{+l}|a_{l m}(\omega_{\rm piv})|^2 .
\end{equation}
In terms of the window function we obtain (see also Refs.~\cite{2010ApJ...710.1089F,Kalashev:2018bra}),
\begin{align}
C_l(\omega_{\rm piv})&=\omega_{\rm piv}^2\int \frac{d\omega_1}{\omega_1}\epsilon(\omega_1)\int dz_1 W\left[\omega_1(1+z_1),z_1\right]\nonumber\\
&\times\omega_{\rm piv}^2\int \frac{d\omega_2}{\omega_2}\epsilon(\omega_2) \int dz_2 W\left[\omega_2(1+z_2),z_2\right]\nonumber\\&
\times\frac{2}{\pi}\int dk k^2 P_\delta\left[k;r(z_1),r(z_2)\right] j_l[kr(z_1)]j_l[kr(z_2)]
\end{align}
where $r(z)=\int_0^z dz/H(z)$ is the comoving distance, and $j_l (kr(z))$ is the spherical Bessel function. 
The power spectrum is defined as $\langle\delta_{\mathbf{k}_1}(r(z_1))\delta_{\mathbf{k}_2}(r(z_2))\rangle=(2\pi)^3\delta^{(3)}(\mathbf{k}_1-\mathbf{k}_2)P_\delta[k_1,r(z_1),r(z_2)]$.
Since the power spectrum varies slowly with $k$, we can apply the Limber approximation~\cite{Ando:2005xg,LoVerde:2008re},
\begin{align}
&\frac{2}{\pi}\int dk k^2 P_\delta[k;r(z_1),r(z_2)]j_l[kr(z_1)]j_l[kr(z_2)]\nonumber\\
&\simeq\frac{1}{r(z_1)^2}  P_\delta\left[k=\frac{l}{r(z_1)};r(z_1),r(z_2)\right]\delta^{(1)}[r(z_1)-r(z_2)].
\end{align}
We are now able to find the correlation over each multipole moment due to the decay of relics,
\begin{align}
C_l(\omega_{\rm piv})&=\int_{0}^{\infty}dz\left[\frac{1}{4\pi}\frac{\omega_{\rm piv}^2 \, 
}{\omega_{\rm max} H(z)}\frac{\rho_{a}}{m_a} \zeta \Gamma_{a\to\gamma\gamma} \right]^2 \nonumber\\
&\times  \left[\epsilon \!\left(\frac{\omega_{\rm max}}{1+z}\right)\right]^2\frac{H(z)}{r(z)^2}  P_\delta \left[k=\frac{l}{r(z)}, r(z),r(z)\right],
\label{eq:Cell}
\end{align}
where the abundance $\rho_a$ and the non-linear spatial power spectrum $P_\delta$ depend on the production mechanism. We  expect the constraints on $\Gamma_{a\rightarrow \gamma\gamma}$ to be the most stringent when $m_a\simeq 2\omega_{\rm piv}$, with a sharp cutoff at smaller masses. At larger masses, the bounds get weaker, since the number density $\rho_a/m_a$ becomes smaller and for a given $\omega_{\rm piv}$ the integral is dominated by the redshift $z\simeq m_a/2\omega_{\rm piv}$.

\section{Production mechanisms}
\label{sec:production}
As we have seen in the previous section, the angular power spectrum contribution from a decaying Big Bang relic depends crucially on its abundance $\rho_a$ and the non-linear spatial power spectrum $P_\delta$ [see Eq.~\eqref{eq:Cell}]. In this section, we will explore a variety of production mechanisms, which will give rise to different abundances and power spectra.

\subsection{Misalignment mechanism and decay of topological defects}
As a first application, we will assume that blue axions constitute 100\% of dark matter. For masses in the range we consider ($5-25\,\rm eV$), freeze-out is not a viable mechanism to produce the entirety of dark matter. Nevertheless, it is well known that the QCD axion~\cite{Peccei:1977hh,Weinberg:1977ma,Wilczek:1977pj}, the pseudo-Nambu-Goldstone boson associated with the spontaneous breaking of the global $U(1)_{\rm PQ}$ Peccei-Quinn symmetry introduced to solve the strong CP problem, is a good candidate for CDM, since it can be produced through the misalignment mechanism~\cite{Preskill:1982cy,Abbott:1982af,Dine:1982ah}. The potential for the field $\phi=|\phi|e^{i\theta}$ includes the terms
\begin{align}
\mathcal{L}&=\frac{\lambda}{4}\left(|\phi|^2-\eta^2\right)^2+\frac{m_a^2\eta^2}{N_{\rm DW}^2}(1-\cos{N_{\rm DW }\theta})\nonumber \\
&-\epsilon_b \lambda^4\frac{|\phi|}{\eta}\cos{(\theta-\delta)},
\end{align}
where $m_a\propto \Lambda_{\rm QCD}^2\eta^{-1}$, and we introduced the bias term $\propto \epsilon_{b}$~\cite{Sikivie:1982qv,Gelmini:1988sf}, possibly related to Planck-suppressed operators~\cite{Barr:1992qq,Kamionkowski:1992mf}, to make the model cosmologically viable if $N_{\rm DW}>1$ and $U(1)_{\rm PQ}$ is broken after inflation. Mutatis mutandis, similar considerations can be applied to a pseudoscalar which does not solve the strong CP problem.

Each of the terms dominates at different times as the temperature of the Universe goes down. When the temperature of the Universe is about $\eta\sim N_{\rm DW} f_a$ ($f_a$ being the axion decay constant), with $N_{\rm DW}$ being the number of minima along the orbit of vacua, cosmic strings form and the phase assumes a value $\theta_0$, the initial misalignment angle. Once the Compton wavelength of the particle $a=\theta \eta$ enters the horizon, the field starts oscillating with frequency $m_a$, producing a QCD axion energy density corresponding to CDM. Notice that the mass depends on temperature, $m_a(T)$, due to the nonperturbative effect of QCD.
If the $U(1)_{\rm PQ}$ symmetry is broken before or during inflation, the QCD axion abundance depends on $\theta_0$. If the symmetry is broken after inflation, axions can be  produced in the decay of cosmic strings~\cite{Harari:1987ht,Hagmann:1990mj} (see Ref.~\cite{Buschmann:2021sdq} and references therein for recent works) and, depending on the UV completion, namely if $N_{\rm DW}>1$, in the annihilation of walls bounded by strings (see e.g.~\cite{Kawasaki:2014sqa}). The latter happens thanks to the bias term, which is needed in order for the $N_{\rm DW}$ post-inflationary scenario to be cosmologically viable, since otherwise the string-wall network would come to dominate the Universe.

Interestingly, the $N_{\rm DW}=1$ post-inflationary scenario predicts a rather small allowed range for the mass of the QCD axion. The misalignment contribution to the abundance depends only on the QCD axion mass (in turn inversely proportional to the $U(1)_{\rm PQ}$ symmetry breaking scale), since $\theta_0$ is averaged over all its possible values. However, a very precise estimate of the contribution from cosmic strings is needed to predict the QCD axion mass as dark matter in the post-inflationary scenario, though (see e.g. Refs.~\cite{Buschmann:2021sdq,Gorghetto:2020qws}). The $N_{\rm DW}>1$ scenario is possible only tuning the value of $\delta$~\cite{Hiramatsu:2012sc,Kawasaki:2014sqa}.

In the  $5-25\,\rm eV$ mass range the QCD axion is largely excluded by several bounds, including the evolution of HB stars in globular clusters, since the coupling to photons is $g_{a\gamma\gamma}=-\frac{\alpha}{2\pi f_a} C_{\alpha\gamma}$, with $C_{\alpha\gamma}=\mathcal{O}(1)$ a model dependent number and $\alpha$ the
fine structure constant. A similar bound, limiting the QCD axion mass to $m_a\lesssim 10^{-1}\,\rm eV$, arises from the constraints on the neutron electric dipole moment~\cite{Graham:2013gfa,Caloni:2022uya}, the only model-independent bound as other couplings can be smaller at the cost of some fine tuning~\cite{Lucente:2022vuo}. Notice however that suppressing the coupling to neutrons (that limits the QCD axion mass to be $m_a\lesssim 10^{-2}\,\rm eV$~\cite{Buschmann:2021juv}) would require even further non-trivial assumptions~\cite{DiLuzio:2017ogq}.
Intriguingly, a heavier-than-expected axion (i.e., with the QCD axion band moved to the right) can result from the existence of $\mathcal{N}$ degenerate Standard Model (SM) replicas, with the axion being the same particle in all the replicas~\cite{DiLuzio:2021pxd}, a generalization of the scenario proposed in Ref.~\cite{Giannotti:2005eb} (see also \cite{DiLuzio:2020wdo}). In this case, the axion mass gets heavier by a factor $\sqrt{\mathcal{N}}$. Therefore, for the blue axion to be the QCD axion, one would need $\mathcal{N}=\mathcal{O}(1000)$ copies of the SM. A more appealing scenario with a heavier-than-expected QCD axion might rely on a single mirror world with a QCD$'$ dynamical scale much larger than the SM $\Lambda_{\rm QCD}$~\cite{Rubakov:1997vp}.
(Other strategies to enlarge the parameter space for the QCD axion can be found in Ref.~\cite{DiLuzio:2021pxd} and references therein.)

Since a rather contrived scenario would be needed for the blue axion to be the QCD axion, one could give up the possibility of solving the strong CP problem, and assume the blue axion to be the pseudo-Nambu-Goldstone boson of a global $U(1)$ symmetry whose breaking scale is unrelated to the coupling to photons $g_{a\gamma\gamma}$ and to the mass $m_a$. Under these assumptions, one can produce the correct DM abundance through misalignment mechanism and cosmic string decay~\cite{OHare:2021zrq}. Moreover, if the orbit of vacua after the $U(1)$ symmetry breaking admits multiple minima, the annihilation of walls bounded by strings due to a small bias in the energy density between the true vacuum and the other minima can easily dominate the production~\cite{Gelmini:2021yzu}.
In this case, the blue axion could be associated with the production of supermassive black hole seeds~\cite{Gelmini:2022nim}. Notice that depending on how large the bias term is (i.e., depending on the annihilation temperature $T_{\rm ann}$ of the string-wall network), blue axions would form potentially at temperatures below $1\rm\, keV$. In such case the matter power spectrum can be affected, and become similar to a warm (or, for even smaller $T_{\rm ann}$, hot) dark matter power spectrum. Since the string-wall network annihilation happens at $T_{\rm ann}\sim 10^{11} \,\mathrm{keV} \sqrt{\epsilon_b m_a/\rm eV}$~\cite{Gelmini:2021yzu,Gelmini:2022nim}, blue axions from the decay of the string-wall network will have a CDM power spectrum if $10^{-22}\ll\epsilon_b\ll 1$. While these figures might seem very unnatural, such small biases can arise from Planck-suppressed operators. Thus, one can produce a power spectrum with a cutoff corresponding to the redshift of the string-wall network annihilation, similar to other late-forming dark matter scenarios~\cite{Das:2020nwc}.

To summarize, we identify the misalignment mechanism and, in the post-inflationary scenario, the decay of cosmic strings and domain walls (the latter for a large enough bias term) as the mechanisms to produce the entirety of dark matter in the form of blue axions, $\rho_a=\rho_{\rm CDM}=\Omega_{\rm CDM}\rho_c$, where $\Omega_{\rm CDM}=0.12 \, h^{-2}$ and $\rho_c=1.05\times 10^{-5} h^2 \, \rm GeV\, cm^{-3}$. 
In this case, the non-linear power spectrum $P_\delta (z, r,r)$ is evaluated with the CLASS code~\cite{Blas:2011rf}, publicly available at~\cite{CLAS}, from redshift $z=0$ to $z=12$ with steps of 0.1. If the bias term is small, the production is dominated by the annihilation of the string-wall network. In this case, we will assume $\rho_a=\rho_{\rm CDM}=\Omega_{\rm CDM}\rho_c$, and $P_\delta (z, r,r)$ equal to the CDM power spectrum, with a cutoff at the co-moving wave-number $k_{T}=7 h \rm \,Mpc^{-1}$, marginally consistent with Lyman-$\alpha$ forest and galaxy clustering data~\cite{Sarkar:2014bca,Das:2020nwc, Gelmini:2022nim}.\footnote{Additional limits might apply, coming from Milky Way satellite analyses~\cite{Das:2020nwc} and from the fraction of primordial isocurvature perturbations~\cite{Gorghetto:2022ikz}, potentially requiring an even larger annihilation temperature. Therefore, our constraints are to be interpreted as conservative, since the larger the annihilation temperature, the closer the power spectrum produced by the string-wall network could get to a CDM power spectrum.}

\subsection{Alternative scenarios for non-cold Dark Matter}

Axions can be copiously produced in the early Universe plasma, and behave as a NCDM component. In this case, the abundance is set by their different interactions, which depend on the considered model~\cite{Turner:1986tb,Chang:1993gm,Masso:2002np,Hannestad:2005df,Graf:2010tv,Archidiacono:2013cha}. We will focus on the coupling to photons, so that the thermal equilibrium is kept by means of the Primakoff process $\gamma Q\to aQ$, where $Q$ refers to any charged particle. Additional interactions (e.g. to nucleons) can keep the axion in thermal equilibrium at smaller temperatures~\cite{Hannestad:2005df,Archidiacono:2013cha,Caloni:2022uya}.

When the interaction rate $\Gamma$ between axions and the SM particles is much larger than the expansion rate $H$, axions are in thermal equilibrium, and eventually decouple from the thermal bath when $\Gamma/H<1$. Therefore, assuming the freeze-out to be instantaneous, decoupling happens at the so-called freeze-out temperature $T_F$ at which the condition $H(T_F) \simeq \Gamma(T_F)$ is satisfied. By requiring $\Gamma \simeq H$, the freeze-out temperature for the Primakoff process is approximately given by~\cite{Cadamuro:2012rm}
\begin{equation}
T_{F}\simeq 4\times10^{4}\GeV\left(\frac{g_{a\gamma\gamma}}{10^{-11}\GeV^{-1}}\right)^{-2}.
\end{equation}
We notice that for $g_{a\gamma\gamma}\lesssim 10^{-9}$~GeV$^{-1}$, $T_F$ is larger than the Electroweak (EW) scale $\Lambda_{\rm EW}$, above which the particle content of the plasma is speculative.

After decoupling, the axion population established at the freeze-out temperature redshifts until today, with abundance~\cite{Kolb:1981hk,Cadamuro:2012rm}
\begin{equation}
    \Omega_{a}h^{2}=\frac{m_{a}}{13~{\rm eV}}\frac{1}{g_{*,s}(T_{F})},
    \label{eq:omega}
\end{equation}
and temperature $T_a$ given by
\begin{equation}
    \frac{T_{a}}{T_{0}}=\left(\frac{g_{*,s}(T_{0})}{g_{*,s}(T_{F})}\right)^{1/3},
    \label{eq:temp}
\end{equation}
where $g_{*,s}(T)$ is the number of entropy-degrees of freedom~\cite{Husdal:2016haj}, $T_{0}=0.235$~meV is the temperature of the Cosmic Microwave Background (CMB), at which $g_{*,s}(T_0)=3.91$. Above the EW scale, the SM provides $g_{*,s}(T_F>\Lambda_{\rm EW})=106.75$ degrees of freedom. Thus, from Eq.~\eqref{eq:omega}, axions with mass $m_a \simeq 155$~eV decoupling at $T_F\gtrsim\Lambda_{\rm EW}$ would account for all the dark matter of the Universe since $\Omega_a\,h^2=\Omega_{\rm CDM}\,h^2 = 0.12$. This implies that, assuming a large reheating temperature, larger masses are excluded~\cite{Cadamuro:2012rm}.

\begin{figure}
	\includegraphics[width=0.45\textwidth]{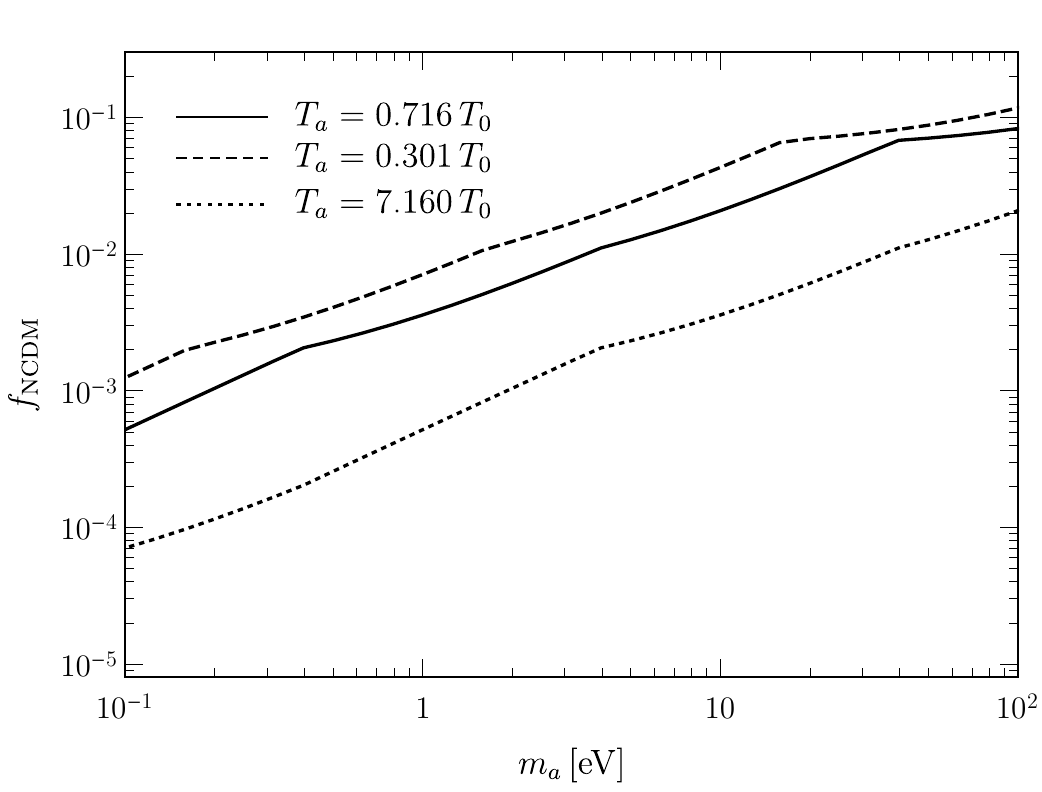}
	\caption{$2\,\sigma$ upper bound on the fraction $f_{\rm NCDM}$ of NCDM as a function of the NCDM axion mass taken from Ref.~\cite{Diamanti:2017xfo} for $T_a = T_\nu = 0.716 T_0$ (solid line), and rescaled for $T_a < T_\nu$ (dashed) and $T_a > T_\nu$ (dotted).}
	\label{fig:fncdm}
\end{figure}

However, the abundance of NCDM axions is constrained to be much smaller than the one of CDM by structure formation observations~\cite{Diamanti:2017xfo}. A NCDM component has a significant free-streaming length, modifying the
matter power spectrum on the smallest spatial scales and affecting the observations of the local Universe. In this context, assuming that the temperature of the NCDM component is the same of the standard neutrinos $T_{\rm NCDM}=T_{\nu}=0.716\,T_0$ and combining the prediction of the number of satellites galaxy with the CMB temperature, polarization and lensing measurements
from the Planck satellite and the Baryon Acoustic Oscillation (BAO) data, the authors of Ref.~\cite{Diamanti:2017xfo} constrained the fraction 
\begin{equation}
f_{\rm NCDM} = \frac{\Omega_{\rm NCDM}}{\Omega_{\rm NCDM}+\Omega_{\rm CDM}}
\label{eq:fncdm}
\end{equation}
of the NCDM component with respect to
the total DM as a function of its mass in the range $(10^{-5}-10^{5})$~eV~\cite{Diamanti:2017xfo}. As shown by the solid line in Fig.~\ref{fig:fncdm} (obtained interpolating data from Fig.~5 of Ref.~\cite{Diamanti:2017xfo}), for $m_a\simeq\mathcal{O}(10)$~eV the $2\,\sigma$ constraint on $f_{\rm NCDM}$ requires $f_{\rm NCDM}\lesssim \mathcal{O}(10^{-2})$, implying $\Omega_{\rm NCDM}h^{2}=1.3\times10^{-3}$ for $m_{a}=4$~eV and $\Omega_{\rm NCDM}h^{2}=8.3\times10^{-3}$ for $m_{a}=40$~eV. Since the cosmological calculations depend on the ratio $m_{\rm NCDM}/T_{\rm NCDM}$, the bound found in Ref.~\cite{Diamanti:2017xfo} can be translated for a model where the NCDM axion has mass $m_a$ and temperature $T_a$ rescaling the mass accordingly, i.e. 
\begin{equation}
\frac{m_a}{T_a} = \frac{m_{\rm NCDM}}{T_{\nu}}.
\label{eq:mresc}
\end{equation}
This implies that the structure formation bound on $f_{\rm NCDM}$ is a function of $m_{\rm NCDM} = m_a\,T_\nu/T_a$. Thus, for $T_a\lesssim T_\nu$ it is shifted to lower masses (see, e.g., the dashed line in Fig.~\ref{fig:fncdm}) and vice versa (dotted line).
From Eq.~\eqref{eq:omega}, for $m_a\sim \mathcal{O}(10)$~eV, the axion relic abundance is much larger than the structure formation constraints~\cite{Diamanti:2017xfo}. This implies that a freeze-out mechanism is ruled out in the context of the SM. We will consider NCDM axions produced in two alternative scenarios:

\begin{itemize}
    \item a modified freeze-out scenario, in which we assume additional degrees of freedom in order to increase $g_{*,s}$, reducing the axion abundance and temperature to satisfy the structure formation bounds;
    \item a hot relic scenario, in which relic axions have the neutrino temperature $T_\nu$ and their abundance saturates the bounds found in Ref.~\cite{Diamanti:2017xfo}.
\end{itemize}

\subsubsection{Modified Freeze-out}
\label{sec:freezeout}
In this scenario, the blue axion decouples at a temperature larger than the EW scale. Since the particle content of the plasma at this high temperature is speculative, there could be additional degrees of freedom increasing the value of $g_{*,s}$. For instance, in the minimal supersymmetric SM scenario, above the supersymmetry breaking energy scale $g_{*,s}=228.75$~\cite{Cadamuro:2012rm}, about twice as much as the SM prediction. From Eqs.~\eqref{eq:omega} and \eqref{eq:temp}, we see that a larger $g_{*,s}(T_F)$ reduces the relic abundance and increases the today NCDM temperature. For an axion with mass $m_a$ and temperature given by Eq.~\eqref{eq:temp}, bounds on the abundance are rescaled following Eq.~\eqref{eq:mresc} and $f_{\rm NCDM}$ is a function of $m_a$ and $g_{*,s}$, since 
\begin{equation}
m_{\rm NCDM} [m_a,\,g_{*,s}(T_F)] = m_a\frac{T_\nu}{T_{0}}\, \left(\frac{g_{*,s}(T_{F})}{g_{*,s}(T_{0})}\right)^{1/3}.
\end{equation}
Thus, from Eq.~\eqref{eq:fncdm},
\begin{equation}
\Omega_a h^2 = \Omega_{\rm CDM} h^2 \frac{f[m_{\rm NCDM}(m_a,\,g_{*,s}(T_F))]}{1-f[m_{\rm NCDM}(m_a,\,g_{*,s}(T_F))]}.
\label{eq:omegancdm}
\end{equation}
Since the abundance of a thermally produced axion is given by Eq.~\eqref{eq:omega}, for a fixed mass $m_a$, equating Eq.~\eqref{eq:omega} to Eq.~\eqref{eq:omegancdm} we find the value of $g_{*,s}$ satisfying the structure formation constraints. Notice that the abundance does not depend strongly on the details of the decoupling, provided that the associated freeze-out temperature is larger than the EW scale. As an example, for $m_a=10$~eV, $g_{*,s}(T_F)=143.479$ is required to satisfy the structure formation bound (independently of the exact value of $T_F > \Lambda_{\rm EW}$), implying a temperature $T_a=0.301\,T_0$. The $2\,\sigma$ upper bound on $f_{\rm NCDM}$ for an axion with $T_a=0.301\,T_0$ is represented by the dashed line in Fig.~\ref{fig:fncdm} (the constraint for $m_a = 10$~eV and $T_a=0.301\,T_0$ is $f_{\rm NCDM}\lesssim 4.3\times 10^{-2}$, equal to the one obtained for a model with temperature $T_a=T_\nu$ and $m_{a}=23.8$~eV, represented by the solid line in Fig.~\ref{fig:fncdm}). For the largest mass considered in this work, $m_a=25$~eV, the structure formation constraints are satisfied by $g_{*,s}(T_F)=199.532$, corresponding to $T_a=0.270\,T_0$.

As discussed in Ref.~\cite{Kalashev:2018bra}, for a thermally produced NCDM axion population, the intensity spectrum is indistinguishable from the one given by decaying CDM. Therefore, for each value of the axion mass, we find the value of $g_{*,s}$ satisfying the structure formation constraints and we evaluate the angular power spectrum using Eq.~\eqref{eq:Cell} with the abundance $\rho_a$ obtained from Eq.~\eqref{eq:omega} and the non-linear NCDM power spectrum $P_\delta$ computed in the adiabatic approximation. In this approximation, if $\mathcal{T}_{\rm NCDM}$ and $\mathcal{T}_{\rm CDM}$ are the transfer functions of NCDM and CDM, respectively, the non-linear NCDM power spectrum is~\cite{Inman:2016qmg}
\begin{equation}
    P_{\delta, {\rm NCDM}} = \left(\frac{\mathcal{T}_{\rm NCDM}}{\mathcal{T}_{\rm CDM}}\right)^{2}P_{\delta,{\rm CDM}}.
    \label{eq:Pdeltancdm}
\end{equation}
Both the transfer functions, relating the primordial and the present day power spectra~\cite{Eisenstein:1997jh}, and the non-linear CDM power spectrum $P_{\delta,\,{\rm CDM}}$ are obtained using CLASS~\cite{Lesgourgues:2011rh}, with
parameters $m_a$, $\Omega_{a}$ from Eq.~\eqref{eq:omega} and $T_{a}$ from Eq.~\eqref{eq:temp}, computed using the value of $g_{*,s}$ satisfying the structure formation bound.

\subsubsection{Hot relic}
\label{sec:relic}
Depending on the presence of additional couplings or particles, blue axions could have a large temperature. We define as ``hot relic'' a particle with the same temperature as neutrinos but with a much larger density, saturating the structure
formation bounds, and stay agnostic about its production mechanism. 
In this scenario, we compute the angular power spectrum $C_l$ using Eq.~\eqref{eq:Cell}, with the axion relic density saturating the structure formation constraint represented by the solid line in Fig.~\ref{fig:fncdm} [without satisfying Eq.~\eqref{eq:omega}] and the non-linear NCDM power spectrum given by Eq.~\eqref{eq:Pdeltancdm}, where $\mathcal{T}_{\rm NCDM}$, $\mathcal{T}_{\rm CDM}$ and $P_{\delta,{\rm CDM}}$ are computed with CLASS with input parameters $m_a$, $T_a = T_\nu= 0.716\,T_0$ and $\Omega_a$ from Ref.~\cite{Diamanti:2017xfo} (the solid line in Fig.~\ref{fig:fncdm}).

We mention here that blue axions could be produced after the decay of heavy dark sector particles, with an energy related to the decaying particle mass. Since the more energetic are the axions, the larger is the effect on structure formation, the relic axion abundance would be more constrained (see the dotted line in Fig.~\ref{fig:fncdm}), while the non-linear power spectrum $P_\delta$ would be suppressed at larger scales. We leave the study of this scenario for future work.

\subsubsection{Freeze-in}
It is possible that axions never reach equilibrium with the thermal bath. Nevertheless, they can still be produced and linger around as dark relics (they ``freeze-in''). We consider that the axion production starts at the reheating temperature $T_{\rm RH}$, when the Universe enters its last phase of radiation domination. For large $T_{\rm RH}$, axions have more time to be produced and their abundance is higher. The lowest value of the reheating temperature compatible with observations is $T_{RH}=5\MeV$~\cite{Kawasaki:2000en,Hannestad:2004px,Ichikawa:2005vw,deSalas:2015glj,Hasegawa:2019jsa,Hasegawa:2020ctq}. 
As further discussed in Ref.~\cite{Langhoff:2022bij}, axions 
with mass $m_a\lesssim T_{\rm RH}$ and interacting with photons are mainly produced via the Primakoff effect and the relic density is given by~\cite{Langhoff:2022bij}
\begin{equation}
    \Omega_{a}h^{2}\simeq10^{-5} \left( \frac{m_a}{\eV}\right)\,\left( \frac{g_{a\gamma\gamma}}{10^{-8}\,\GeV^{-1}}\right)^2 \left(\frac{T_{RH}}{5\,\MeV}\right).
    \label{eq:omegafreezein}
\end{equation}
For the purpose of this work, the freeze-in paradigm will not give an axion relic abundance large enough to have observable effects, so we will not discuss it any further.

\begin{figure}
	\includegraphics[width=0.45\textwidth]{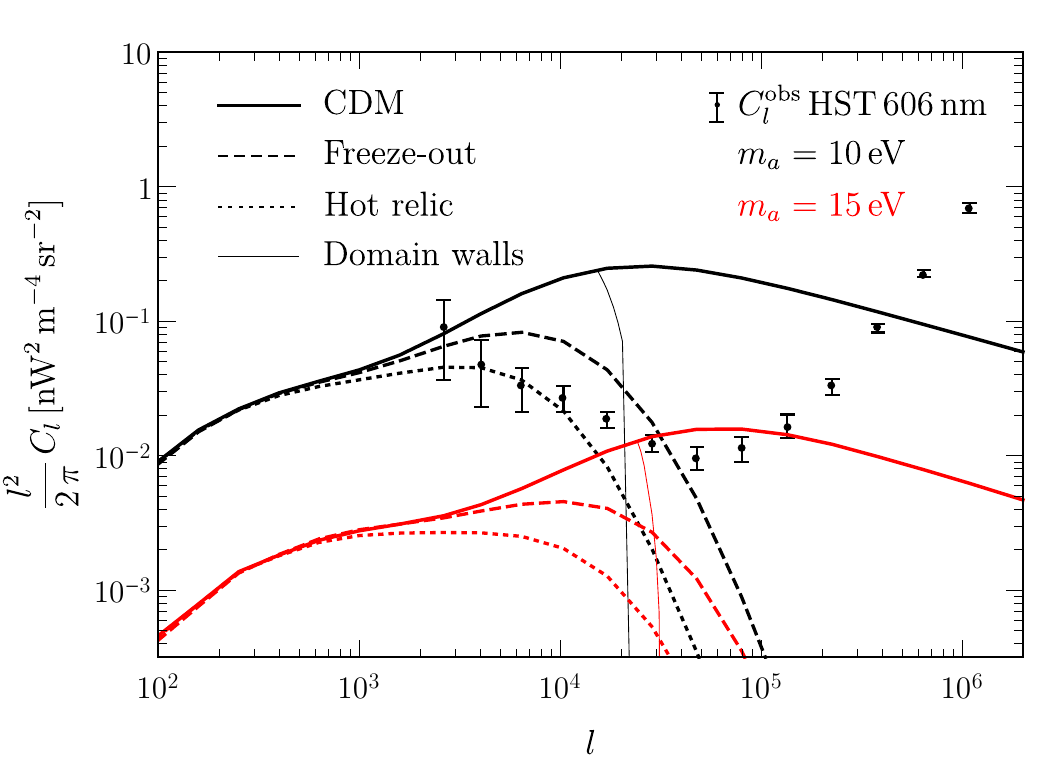}
	\caption{Angular power spectrum of the COB anisotropy for the DM axion decaying into photons, for $\lambda_{\rm obs} = 606$~nm, in the case of CDM (solid lines), freeze-out (dashed), hot relic (dotted) and CDM produced by annihilation of domain walls (thin-solid). We have fixed the axion mass \mbox{$m_{a}=10$~eV} (black lines) and $m_{a}=15$~eV (red) and the product \mbox{$\rho_a\,\Gamma_{a\to\gamma\gamma}=10^{-29}\,\rm s^{-1}\,\GeV\,\cm^{-3}$}. 
    The HST measurements at 606 nm are shown with error bars~\cite{Mitchell-Wynne:2015rha}.}
	\label{fig:clfree}
\end{figure}
\section{Results}
\label{sec:results}
In this section we compare the resulting angular power spectra in the different scenarios previously discussed with the HST measurements in order to constrain the axion parameter space. In particular, we use the dataset at the shortest available wavelength, namely 606 nm, obtained by the Wide Field Camera 3 and the Advanced Camera for Surveys, which covers 120 square arcminutes in the Great Observatories Origins Deep Survey~\cite{Mitchell-Wynne:2015rha}. 

\subsection{Bounds and future reaches on cold dark matter}

\begin{figure}[t!]
	\includegraphics[width=0.45\textwidth]{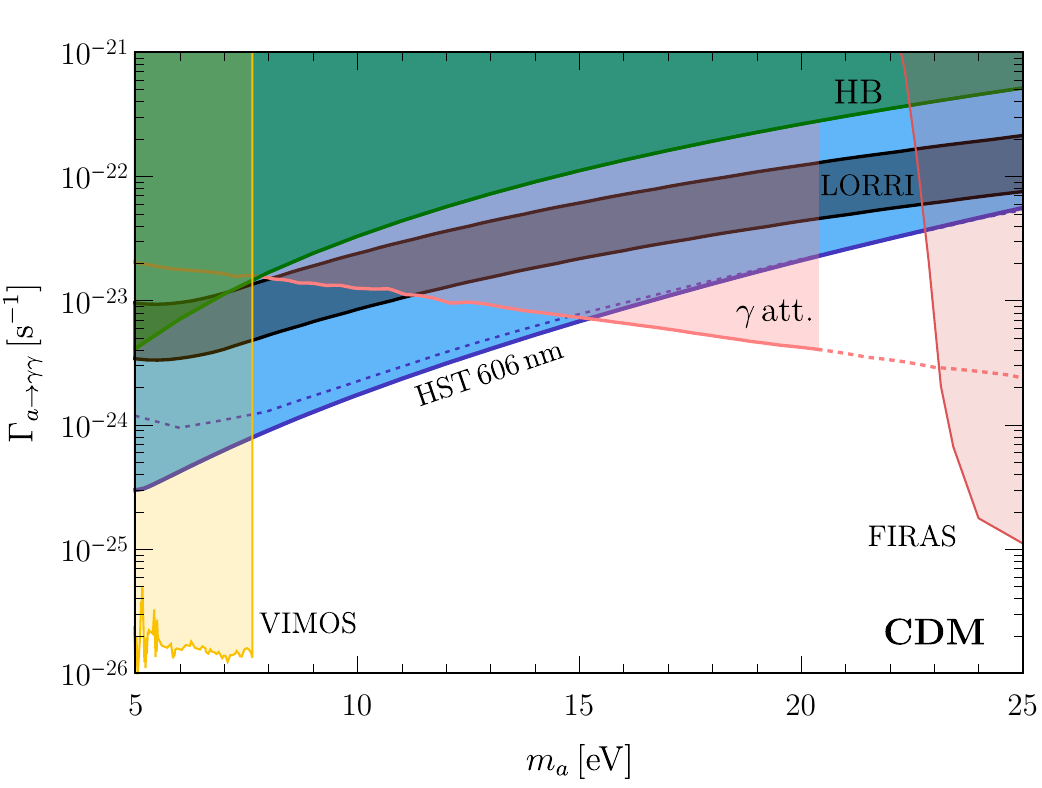}
	\caption{Comparison between our bound on CDM using HST measurements at 606 nm (solid blue line) and the bound obtained in Ref.~\cite{Nakayama:2022jza} (the dotted blue line). We also show other bounds. The color code is the same as Fig.~\ref{fig:boundratecdm}.}
	\label{fig:comparison}
\end{figure}

In Fig.~\ref{fig:clfree} we show the angular power spectrum $l^2\,C_l/2\pi$ as measured by HST at 606 nm with error bars, and the angular power spectrum from decaying CDM as seen by the detector at 606 nm, at fixed value of $\rho_a\times\Gamma_{a\to\gamma\gamma}=10^{-29}$~s$^{-1}$~GeV~cm$^{-3}$ for $m_a=10$~eV (black lines) and $m_a=15$ (red lines). We constrain the blue axion lifetime by requiring that the angular power spectrum $C_l$ in Eq.~\eqref{eq:Cell} does not exceed the upper error bar of any of the data points. In the case of CDM (solid lines) the quantity $l^2\,C_l$ is peaked at $l\gtrsim 10^4$ and the peak is shifted towards smaller scales as the mass increases. In this context, the most constraining bin is the one at $l\simeq 4.8\times 10^5$ and we exclude the light blue region delimited by the solid line and labelled as HST 606 nm in Fig.~\ref{fig:boundratecdm}, finding good agreement with the results in Ref.~\cite{Nakayama:2022jza} for $m_a\gtrsim 10$~eV. In particular, as shown in Fig.~\ref{fig:comparison}, where the bound from Ref.~\cite{Nakayama:2022jza} is represented by the dotted blue line, our bound is stronger by $\sim 30\%$ at $m_a=10$~eV, by $\sim 15\%$ at $m_a=15$~eV and the discrepancy becomes negligible for even larger masses. For smaller values of the mass, we find a larger discrepancy between the two results, with our bounds stronger by a factor $\sim 4$ at $m_a\simeq 5$~eV. This could be due to the several differences in our analysis. Differently from Ref.~\cite{Nakayama:2022jza}, we get the non-linear power spectrum directly from CLASS. In addition, we characterize the detector response more precisely, by considering the detector $\omega_{\rm piv}$ and the throughput (see Appendix~\ref{app:HST}) instead of using an observational bandwidth equal to the observational frequency. We checked that using the same detector response as the one used in Ref.~\cite{Nakayama:2022jza}, the discrepancy at lower masses is reduced. 

\begin{figure}
	\includegraphics[width=0.45\textwidth]{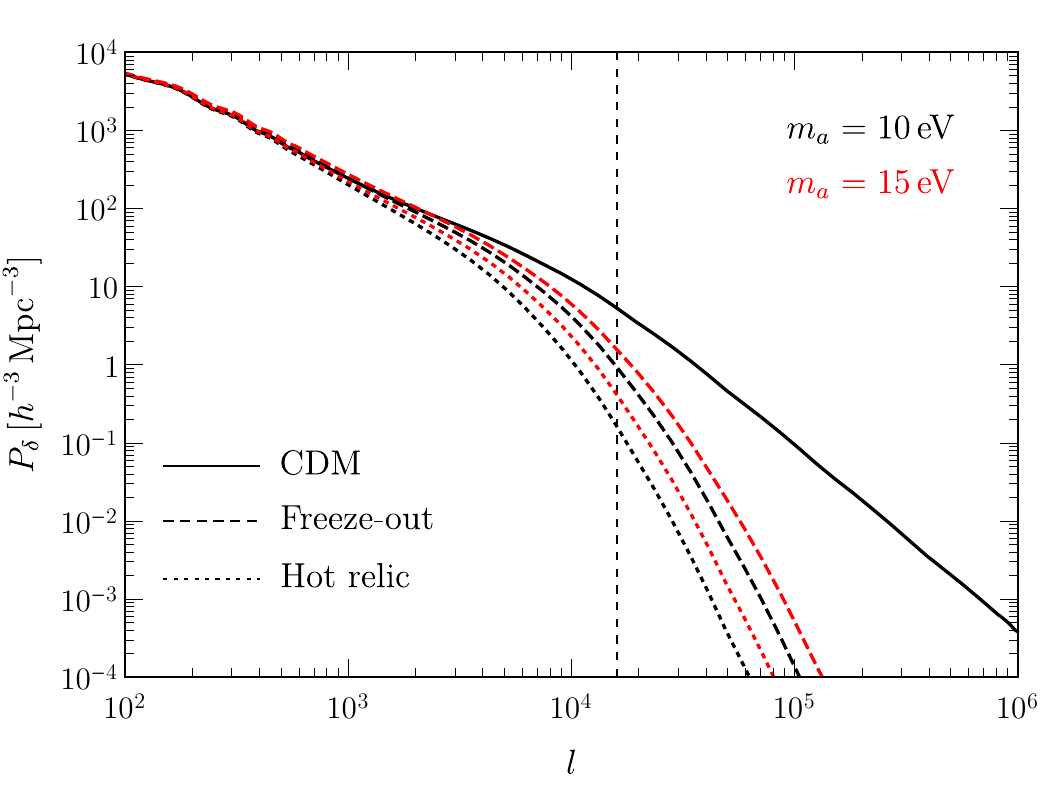}
	\caption{Non-linear spatial power spectrum for CDM (black solid line), freeze-out (dashed) and hot relic (dotted) at redshift $z=1$ for $m_a=10$~eV (black lines) and $m_a=15$~eV (red). The vertical dashed line at $l_T=1.6\times 10^{4}$  corresponds to the cutoff comoving wave-number $k_T = 7\,h\,\Mpc^{-1}$ for CDM produced by annihilation of domain walls. In this latter case, the non-linear power spectrum is assumed to be the one for CDM (solid black line), cut at $l_T$.}
	\label{fig:pnl}
\end{figure}

\begin{figure*}
	\includegraphics[width=0.49\textwidth]{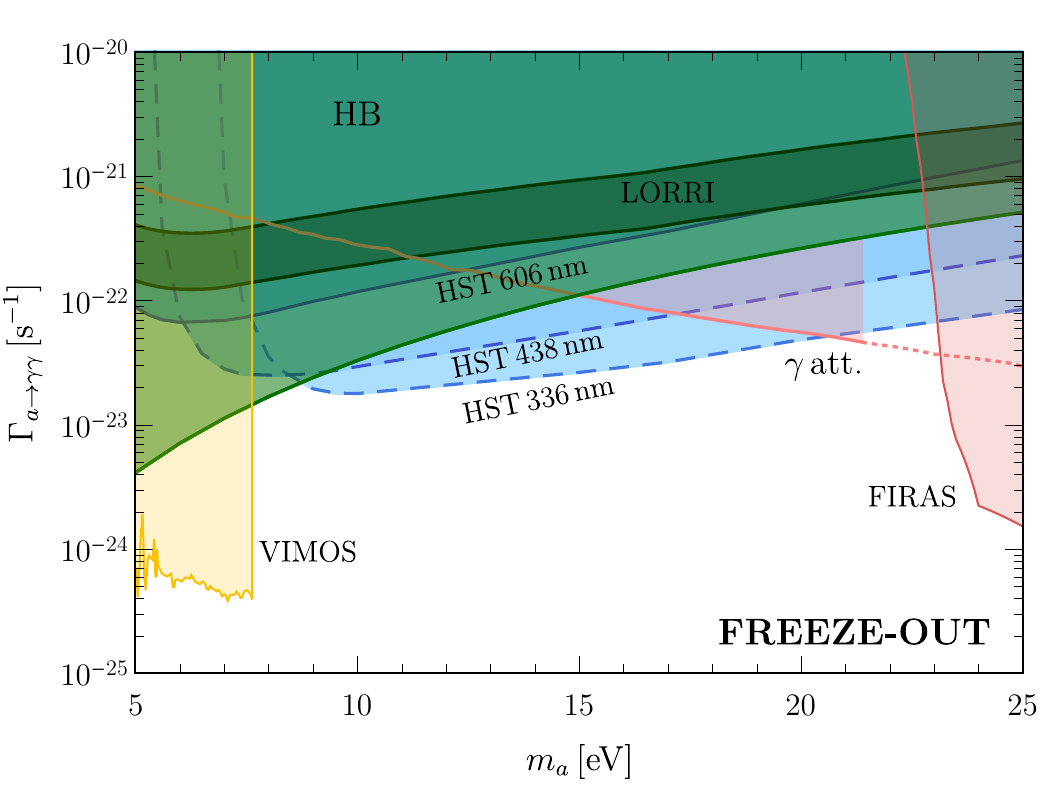}\,
	\includegraphics[width=0.49\textwidth]{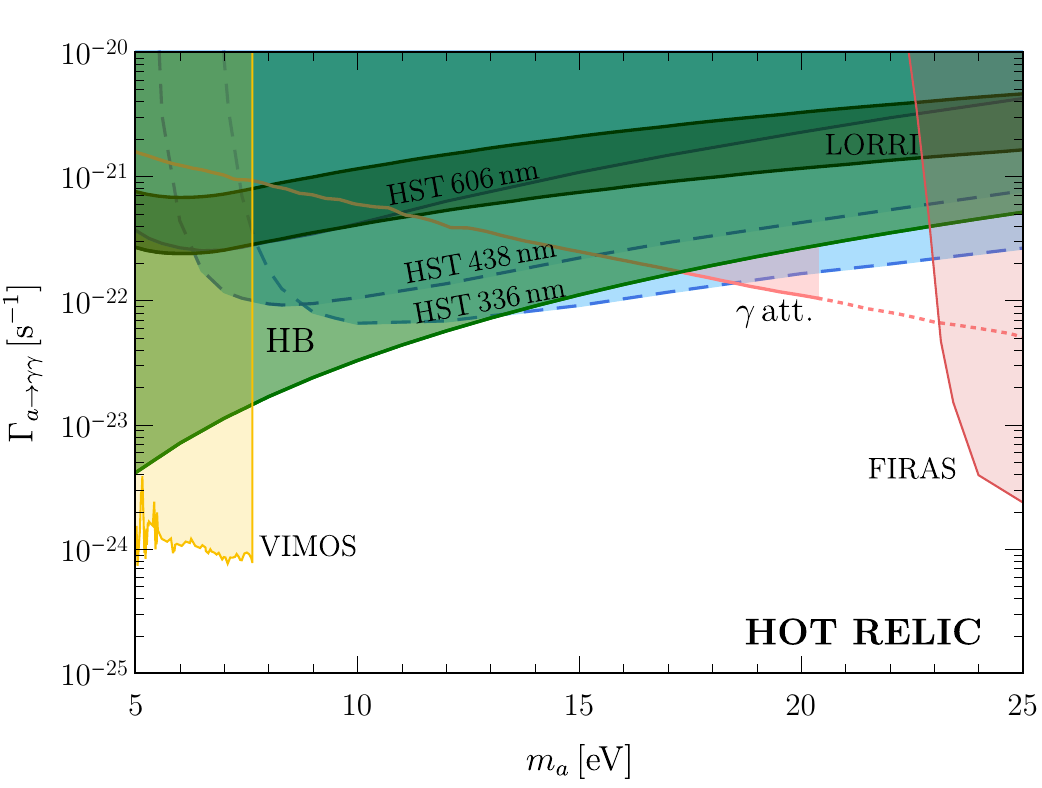}\,
	\includegraphics[width=0.49\textwidth]{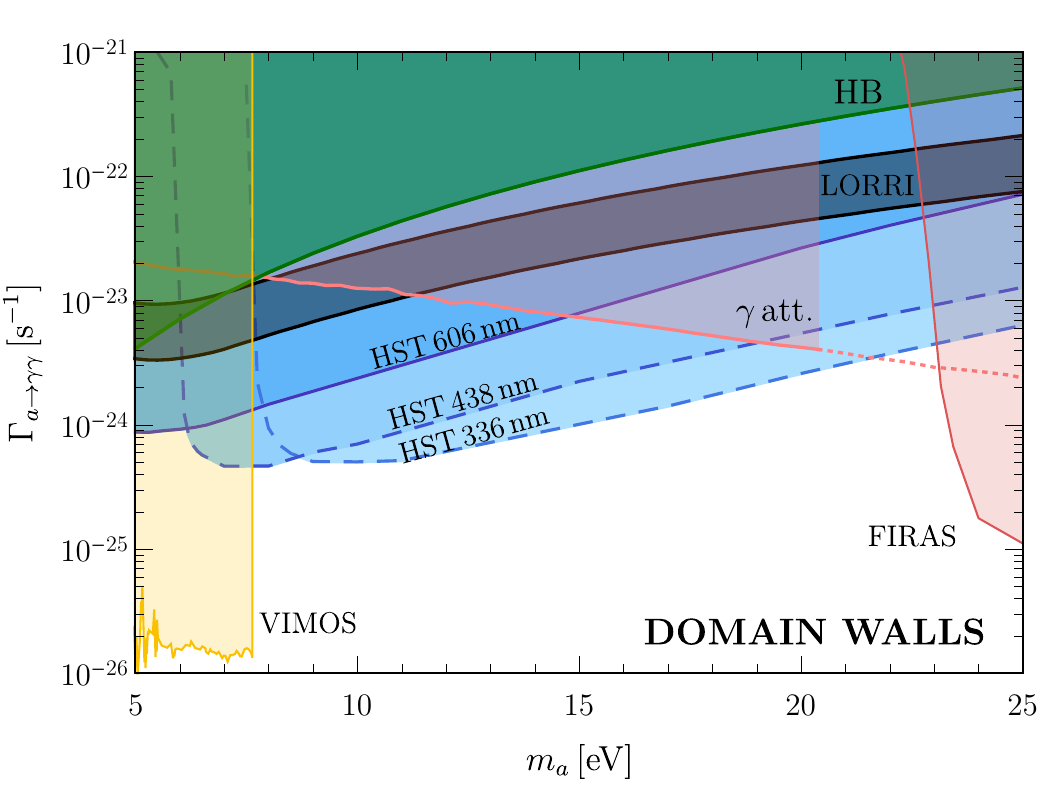}\,
	\caption{Bounds and future reaches on the lifetime of the blue axion in the case of freeze-out (upper left), hot relic (upper right) and CDM produced by annihilation of domain walls (lower panel). We show also other bounds. The color code is the same as Fig.~\ref{fig:boundratecdm}.} 
	\label{fig:boundrate}
\end{figure*}

Stronger constraints on the axion lifetime could be obtained with HST measurements at shorter wavelengths. In order to forecast the possible future reaches, we evaluate the angular power spectrum in Eq.~\eqref{eq:Cell} using $\omega_{\rm piv}$ and $\epsilon(\omega)$ for measurements at 438 nm and 336 nm, respectively (see Appendix~\ref{app:HST} for more details) and we require that it does not exceed the upper error bar of any data points at 606 nm. In this way we probe the light blue regions delimited by dashed lines in Fig.~\ref{fig:boundratecdm}, labelled with HST 438 nm and HST 336 nm, respectively. In the case of 336 nm, the bound would be improved by a factor 4-10 in the mass range $10-25$~eV, leading to the strongest constraints in this region. We stress that our assumptions are conservative, since at smaller wavelengths the measured angular power spectrum is expected to be even smaller, implying even stronger projected reaches.

\subsection{Alternative scenarios}
The anisotropy bounds can be relaxed in alternative scenarios. The resulting angular power spectrum $C_l$ can be smaller due to the reduction in the abundance $\rho_a$ and the suppression of the non-linear power spectrum $P_\delta$, depending on the production mechanism. In Fig.~\ref{fig:clfree} we show the resulting angular power spectrum for the cases of freeze-out (dashed) and of a hot relic with neutrino temperature (dotted). The effect of the abundance suppression is blurred out since the angular power spectra are shown at fixed value of $\rho_a\Gamma_{a\to\gamma\gamma}$ (a larger value of $\Gamma_{a\to\gamma\gamma}$ compensates the suppressed $\rho_a$). On the other hand, the effect of the non-linear power spectrum suppression is clearly visible, since for NCDM the quantity $l^2\,C_l$ is suppressed at $l\lesssim 10^4$. The suppression point is at larger scales for hotter DM and, for a fixed cosmological scenario, the suppression starts at smaller scales as the mass increases. Indeed, this reflects the trend of the non-linear power spectrum, as shown in Fig.~\ref{fig:pnl} at a representative value of the redshift $z=1$ for $m_a=10$~eV (black) and $m_a=15$~eV (red) in the different scenarios that we considered. The most conservative scenario is the hot relic case, since the angular power spectrum is suppressed at larger scales. Again, by requiring that the computed $C_l$ must not exceed the error bar of any data points at 606 nm, we compute the bounds for the observations at 606 nm and forecast the sensitivity for future observations. In Fig.~\ref{fig:boundrate} we show how the bounds on the axion lifetime are relaxed in the freeze-out (upper-left panel) and hot relic (upper-right panel) scenarios (for a discussion on the other bounds and how they change in the different scenarios see Appendix~\ref{app:bounds}). In these scenarios, future measurements at shorter wavelengths would improve the current anisotropy bound by almost one order of magnitude, setting the strongest probe on the blue axion lifetime for $10~\eV \lesssim m_a \lesssim 20~\eV$ (freeze-out) and $15~\eV \lesssim m_a \lesssim 18~\eV$ (hot relic).

Finally, we show what happens in the case in which axions are produced by the annihilation of a string-wall network. We model this case using a CDM $P_\delta$ (the black solid line in Fig.~\ref{fig:pnl}) featuring a cutoff at the co-moving wave-number $k_T=7\,h\,\Mpc^{-1}$, represented by the vertical dashed line in Fig.~\ref{fig:pnl}. As shown by the thin-solid lines in Fig.~\ref{fig:clfree}, in this scenario the resulting angular power spectrum is the same as the CDM case at large scales and it is strongly suppressed at $l\simeq 10^4$, due to the cutoff previously discussed. Therefore, the most constraining data point is at larger scales compared to the CDM case (depending on the axion mass) and the constraints are relaxed by a factor $\sim 2$, as shown in the lower panel of Fig.~\ref{fig:boundrate}. However, also in this case, the HST measurements at 606 nm already exclude the CDM interpretation of the LORRI excess.

\begin{figure*}[t!]
	\includegraphics[width=0.49\textwidth]{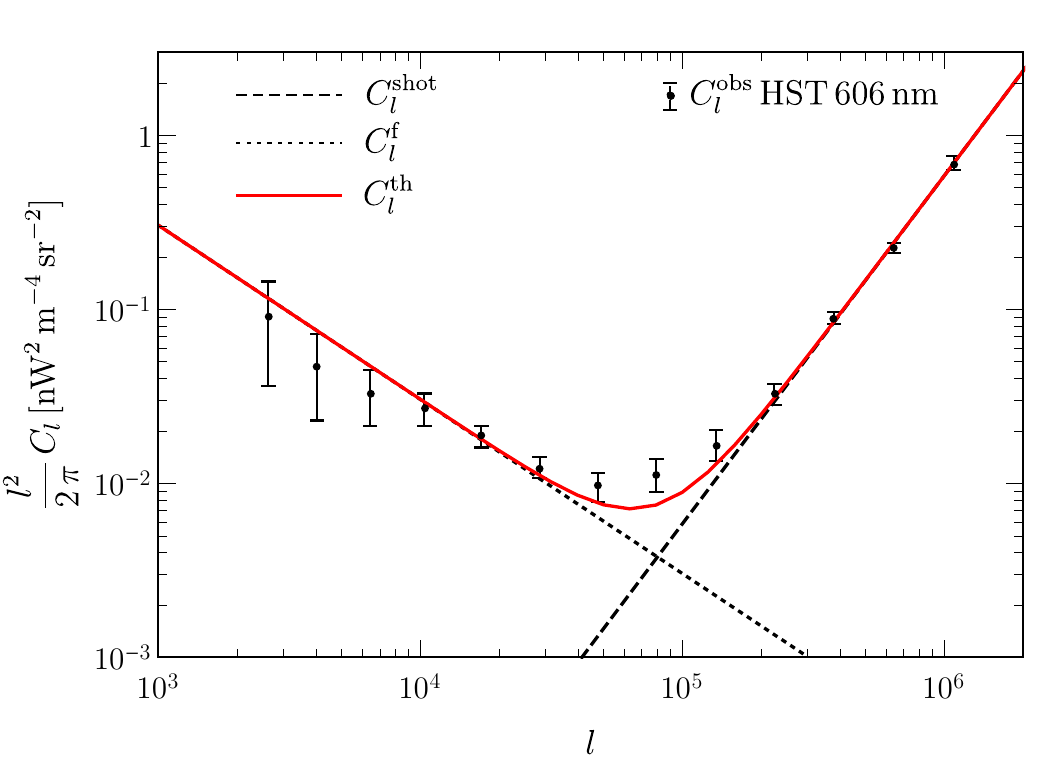}
	\includegraphics[width=0.49\textwidth]{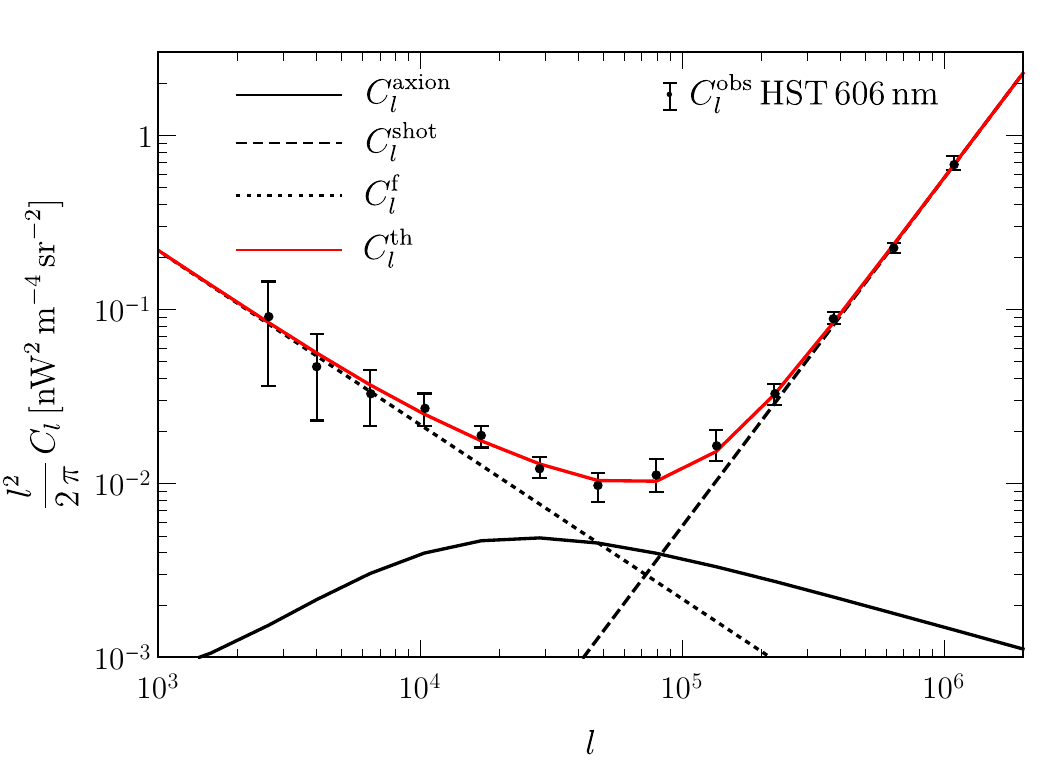}\,
	\caption{Best fit in the absence of axions (left panel) and in the case of CDM axion with $m_a=10$~eV (right panel). In both the panels, we show the HST measurements with error bars, the shot noise (dashed black line), the foreground (dotted black), the axion (solid black) contributions and their sum (red line).} 
	\label{fig:bestfit}
\end{figure*}

\subsection{Model constraints}

\begin{figure*}
	\includegraphics[width=0.49\textwidth]{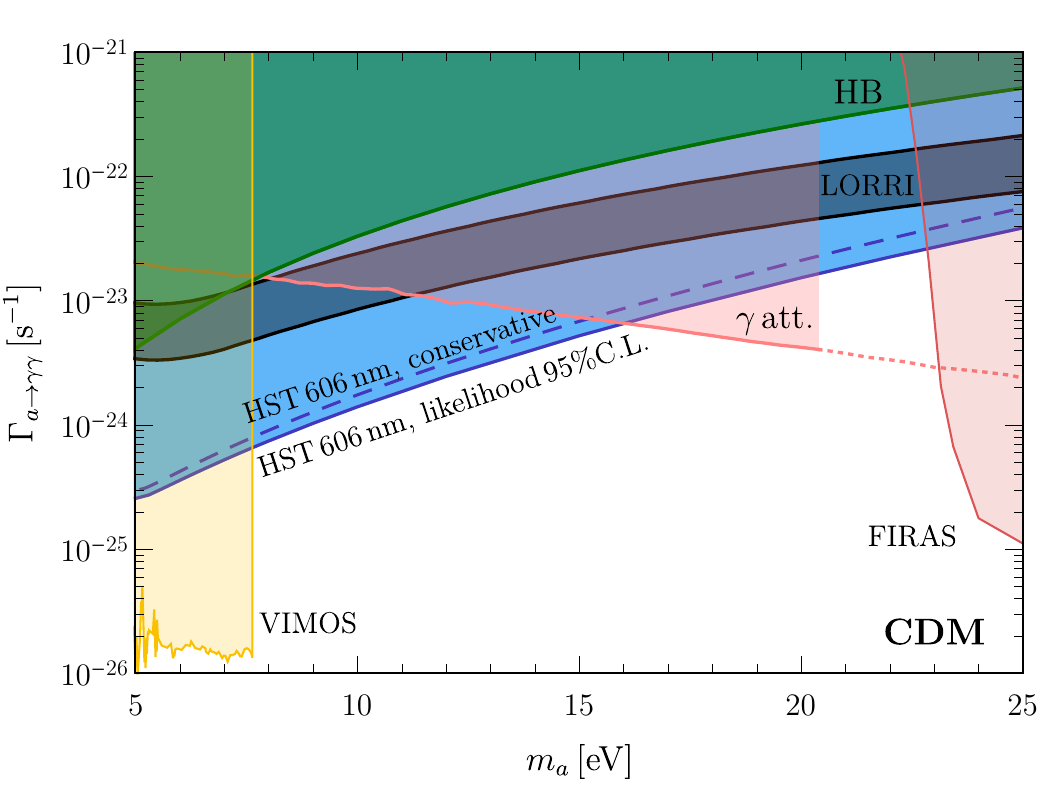}
	\includegraphics[width=0.49\textwidth]{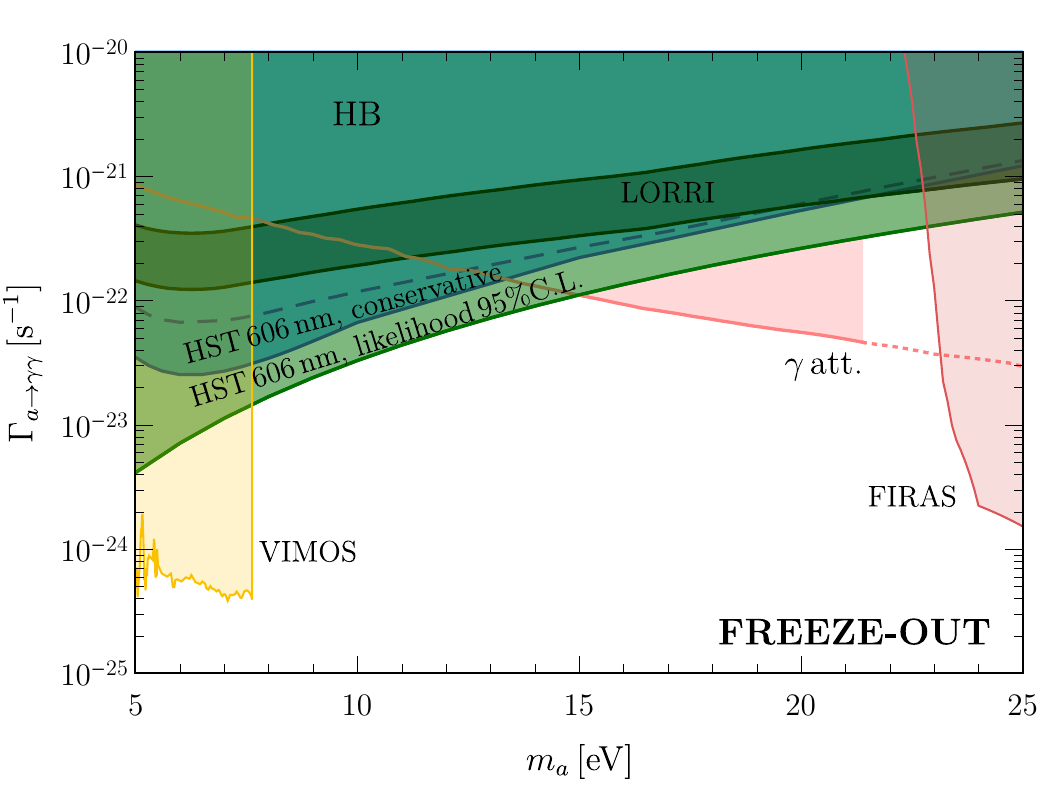}\,
	\includegraphics[width=0.49\textwidth]{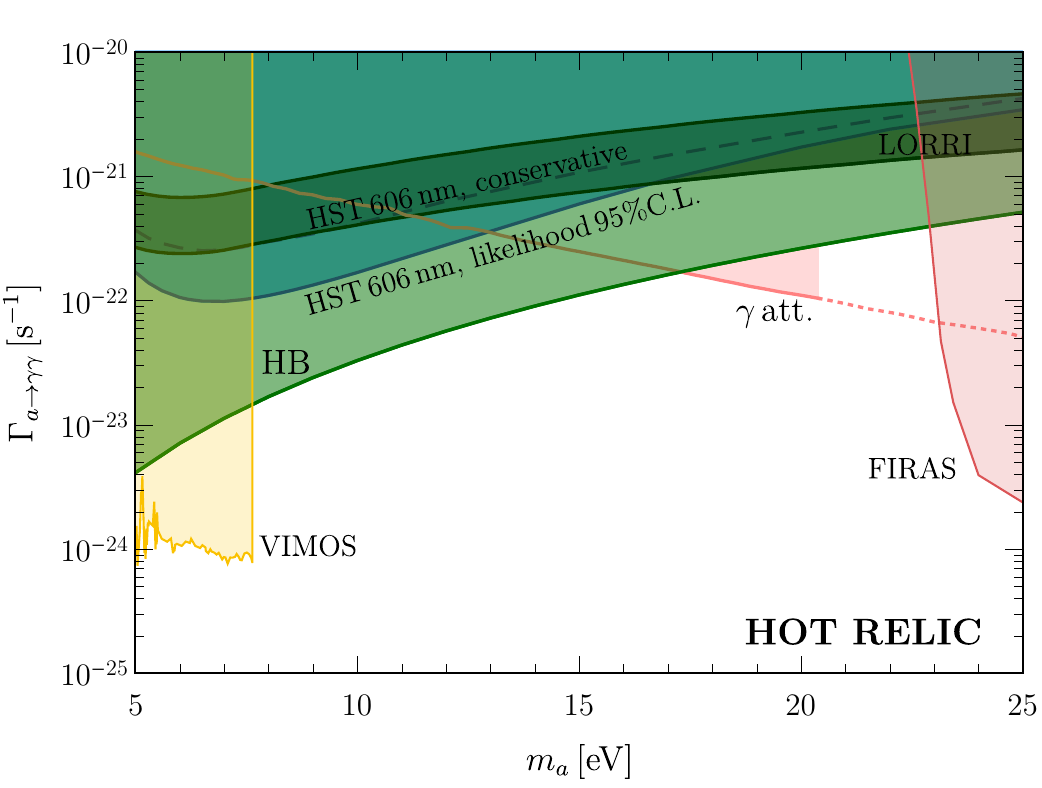}\,
	\includegraphics[width=0.49\textwidth]{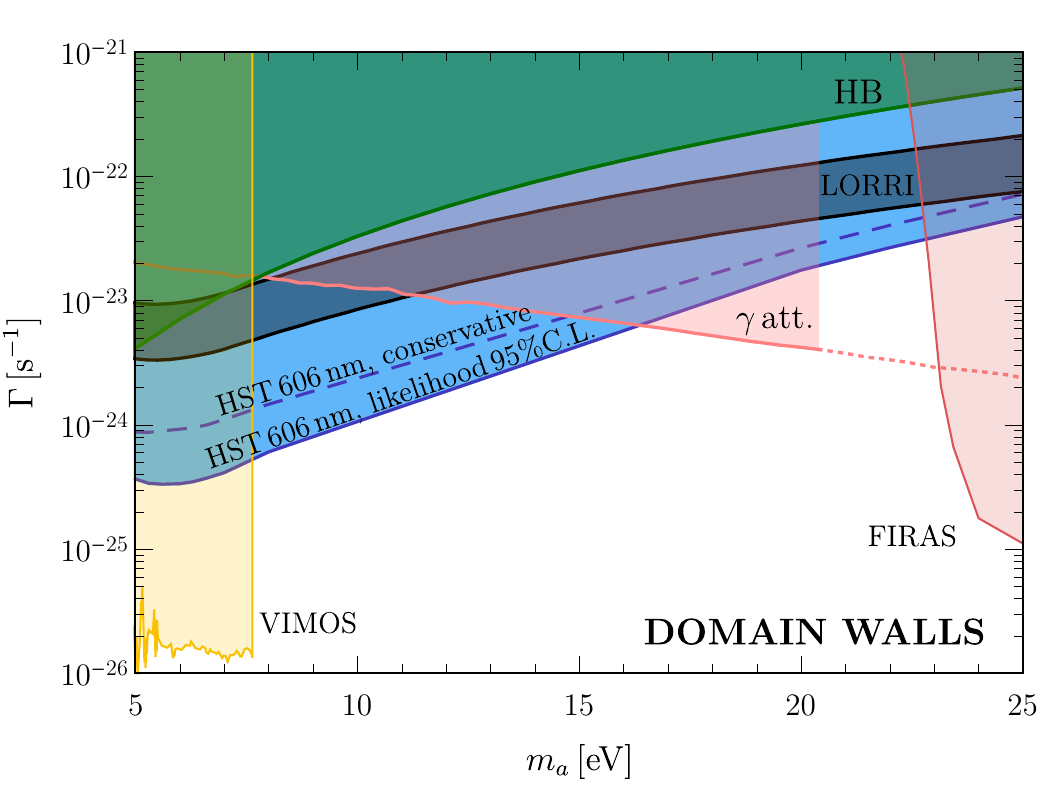}\,
	\caption{Difference between the bound from HST measurements at 606 nm obtained requiring that $C_l$ must not exceed the upper error bar of any data points (conservative, dashed blue line) and the $95\%$ CL constraint from the likelihood analysis (solid blue line) in the case of CDM (upper-left panel), freeze-out (upper right), hot relic (lower left) and CDM produced by annihilation of domain walls (lower-right panel).  We show also the other bounds and the excess in LORRI data with the same color code as Fig.~\ref{fig:boundratecdm}}
	\label{fig:boundchi}
\end{figure*}

Our bounds on the blue axion lifetime are conservative since we require that the blue axion decay contribution saturates the COB anisotropy spectra. Moreover, they are robust, since they parametrically depend on the square root of the angular power spectrum. Stronger constraints can be obtained if we model the HST data through three different components, including the axion decay contribution, the shot-noise term and a power-law component accounting for Galactic foregrounds, e.g. the diffuse Galactic light (DGL) \cite{Gong:2015hke}. In principle, one should consider also the signal from high-z faint galaxies during the
epoch of reionization. However, we neglect it since for the HST measurements at 606 nm there is
almost no contribution from high-$z$ signal~\cite{Mitchell-Wynne:2015rha,Gong:2015hke}.   At small scales, the power spectrum is dominated by the shot noise, which is scale-independent and with an angular power spectrum given by \cite{Gong:2015hke}
\begin{equation}
    C_l^{\rm shot} = A_{\rm shot},
\end{equation}
where $A_{\rm shot}$ is the shot-noise amplitude factor, which is constant for a given observed wavelength. At large scales, the power spectrum is dominated by $C_{l}^{f}$ for foregrounds, which can be described by \cite{Gong:2015hke}
\begin{equation}
    C_l^f = A_f\,l^{-3},
\end{equation}
where $A_f$ is the amplitude factor for the foregrounds. For instance, a possible dominant component in $C_l^f$ could be the DGL, since the DGL term is proportional to $\sim l^{-3}$. Therefore, closely following Ref.~\cite{Gong:2015hke}, the HST data can be fit as
\begin{equation}
\begin{split}
    & C_l^{\rm th}(m_a,\,g_{a\gamma\gamma},\,A_{\rm shot},\,A_{f}) = \\
    & C_l^{\rm axion}(m_a,\,g_{a\gamma\gamma}) + C_l^{\rm shot}(A_{\rm shot}) + C_l^{f} (A_f),
\end{split}
\end{equation}
with $C_l^{\rm axion}$ given by Eq.~\eqref{eq:Cell}.

In order to constrain the axion parameter space, we perform a likelihood analysis, by defining the likelihood function $\mathcal{L} \propto \exp (-\chi^2/2)$, with $\chi^2$ given by
\begin{equation}
\begin{split}
    \chi^2(m_a,\,g_{a\gamma\gamma},\,A_{\rm shot},\,A_{f}) = \sum_{i=1}^{N_d} \frac{\left(C_{l,i}^{\rm obs} - C_{l,i}^{\rm th}\right)^2}{\sigma_i^2},
\end{split}
\label{eq:chisq}
\end{equation}
where $N_d=13$ is the number of data points, $C_{l,i}^{\rm obs}$ is the angular power spectrum from the $i$-th observational point with error $\sigma_i$, and $C_{l,i}^{\rm th}$ is the theoretical angular power spectrum evaluated at $l$ corresponding to the $i$-th data point. We assume the free parameters to follow a flat prior distribution, with $A_{\rm shot}\in (10^{-13}\,,\,10^{-11})$ and $A_{f}\in (10^{2}\,,\,10^{4})$~\cite{Mitchell-Wynne:2015rha}. In the left panel of Fig.~\ref{fig:bestfit} we show the best fit in the absence of axions, with $A_{{\rm shot},0}=3.66\times 10^{-12}$ and $A_{f,0}=1.91\times 10^{3}$. As shown by the value of the chi squared $\chi^2_0 = 9.2$ over 11 degrees of freedom, the fit well reproduces the data points, with the largest discrepancy for $l\sim 10^5$ where data are underestimated.\\
In order to constrain the axion, we marginalize over $A_{\rm shot}$ and $A_f$ by minimizing the $\chi^2$ for each value of $m_a$ and $g_{a\gamma\gamma}$ over $A_{\rm shot}$ and $A_f$,
\begin{equation}
    \bar{\chi}^2 = \min_{A_{\rm shot},\,A_{f}}\chi^2 (m_a,\,g_{a\gamma\gamma},\,A_{\rm shot},\,A_{f})\,. 
\end{equation}
If axions are CDM, for certain values of the coupling, their presence can improve the fit. In this case, as shown in the right panel of Fig.~\ref{fig:bestfit} for $m_a=10$~eV, the quantity $l^2 C_l^{\rm axion}$ is peaked at $l\sim 10^4-10^5$, reducing the discrepancy between data and fit at those scales. For instance, for $m_a=10$~eV, the best fit is obtained for $g_{a\gamma\gamma}=1.24\times 10^{-11}$~GeV$^{-1}$, $A_{\rm shot}=3.56 \times 10^{-12}$ and $A_f=1.28\times 10^3$. However, we mention that this should not be considered as a hint for the blue axion. Indeed, axions improve the fit due to the uncertainty related to the standard physics contributions and including them we overfit the data, obtaining $\chi^2_{\min}=2.02$ over 10 degrees of freedom for $m_a=10$~eV. Moreover, the cross-correlations between intensity fluctuations at different wavelengths due to dark matter decay is zero (as structures at
different redshifts are mainly uncorrelated), contrary
to the observations~\cite{Gong:2015hke}. Therefore, more than one axion (in Ref.~\cite{Gong:2015hke} a continuous distribution of axion masses is assumed) is needed to fit the cross-correlations.

In the case of NCDM, the presence of axions would not improve the fit, since the angular power spectrum is suppressed at larger scales $l\lesssim 10^4$, as shown in Fig.~\ref{fig:clfree}. In this case, the minimum of $\chi^2$ is $\chi_0^2=9.19$ for vanishing axion-photon coupling. We define the test statistic
\begin{subequations}\label{cases}
\begin{empheq}[left={\chi_*^2 =   \empheqlbrace\,}]{align}
 &\bar{\chi}^2 - \bar{\chi}^2_{\rm min}&      g_{a\gamma\gamma}\geq g_{\rm min} \\
  &0&       g_{a\gamma\gamma} < g_{\rm min}
\end{empheq}
\end{subequations}
where $\bar{\chi}^2_{\rm min}$ is the minimum chi squared, obtained at the value of the axion photon coupling $g_{a\gamma\gamma}=g_{\rm min}$ ($g_{\rm min} \neq 0$ only in the CDM case). This quantity follows a half-chi-squared distribution~\cite{Cowan:2010js}, which allows us to constrain the axion parameter space at $95 \%$ Confidence Level (CL) by requiring $\chi_*^2 \leq 2.7$. As shown in the upper-left panel of Fig.~\ref{fig:boundchi}, in the CDM case this approach is consistent with the conservative one previously discussed and it strengthens bounds on the rate by $\sim 15\%$ at $m_a=5$~eV and $\sim 25\%$ at $m_a=25$~eV. In the other cases, this approach leads to factor $\sim 2$ stronger constraints for $m_a\lesssim 10$~eV, with a lower discrepancy at larger masses, as shown in the remaining panels of Fig.~\ref{fig:boundchi}. Indeed, when alternative cosmological scenarios are considered, the angular power spectrum is suppressed at larger scales. Since with our conservative approach we set the bound when the angular power spectrum exceeds any of the data points, the most constraining one is at $l \lesssim 10^4$, where $l^2 C_l$ is larger, without taking into account all the other measurements. On the other hand, by performing a likelihood analysis that includes additional contributions, all the data points are important since by definition the $\chi^2$ accounts for the entire data set,  as shown in Eq.~\eqref{eq:chisq}, allowing us to set stronger bounds.

\section{The Dark Portal}
\label{sec:dark}
\begin{figure*}
	\includegraphics[width=0.49\textwidth]{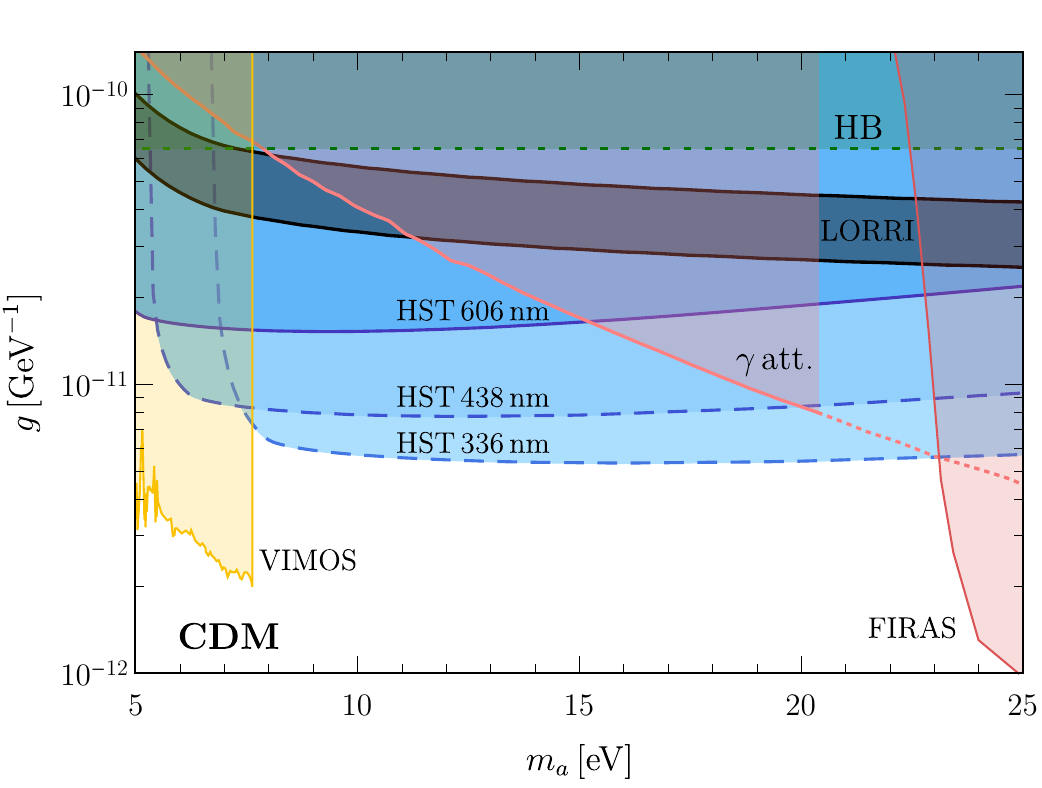}
	\includegraphics[width=0.49\textwidth]{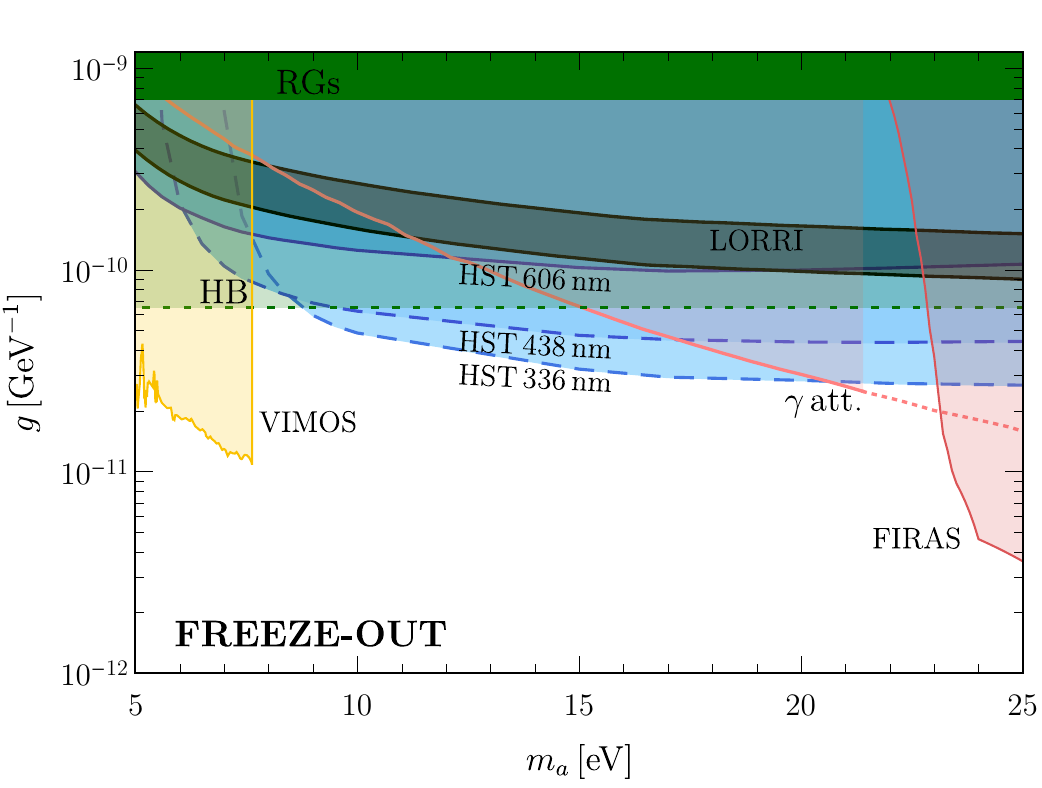}\,
	\includegraphics[width=0.49\textwidth]{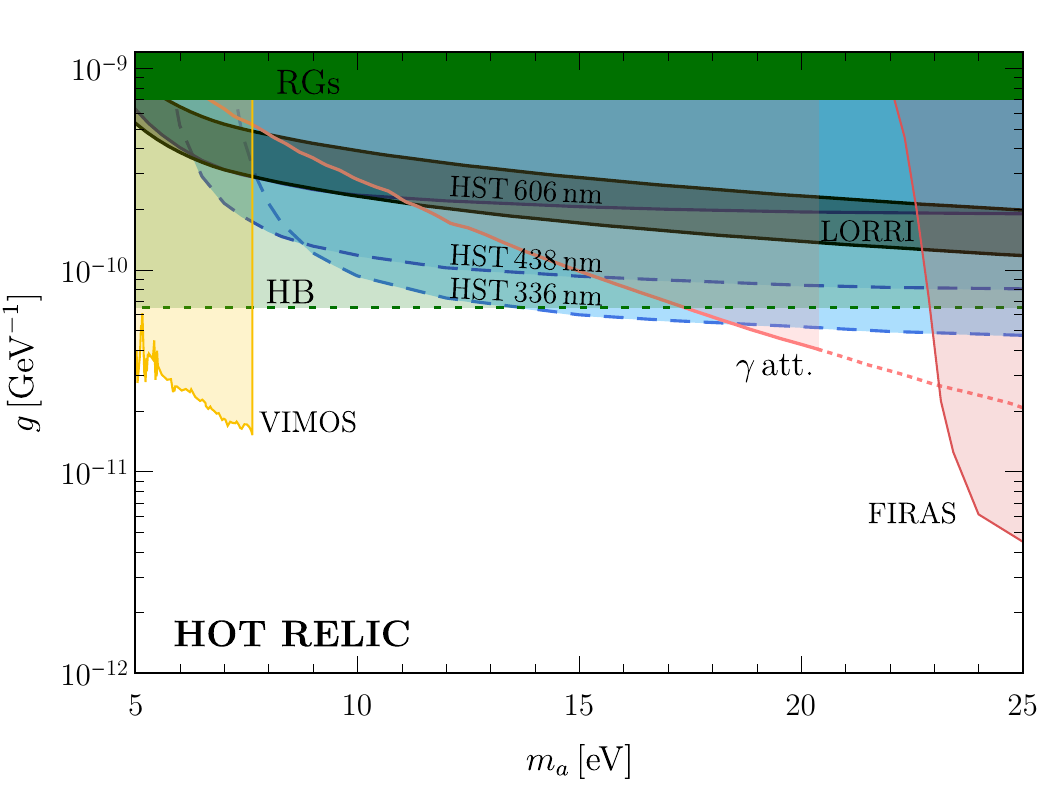}\,
	\includegraphics[width=0.49\textwidth]{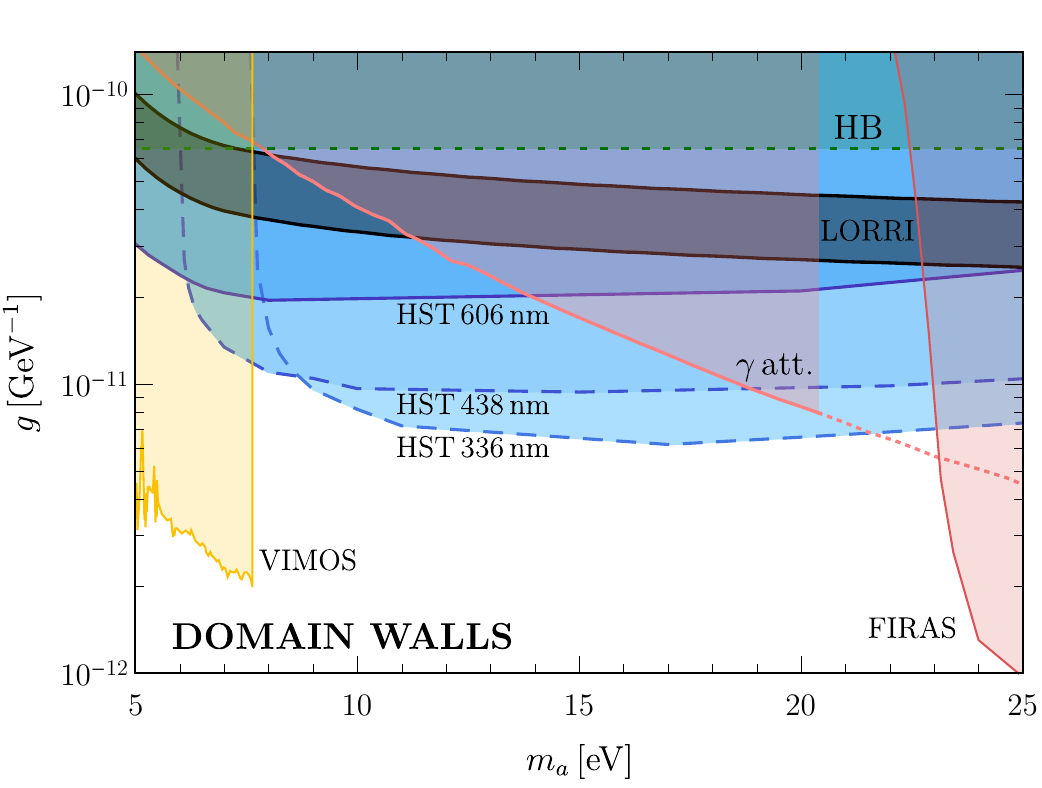}\,
	\caption{Bounds and future reaches on the coupling $g$ of the blue-axion in the case of CDM (upper-left), freeze-out (upper-right), hot relic (lower-left) and late-time produced CDM produced by annihilation of domain walls (lower-right panel). If $g=g_{a\gamma\gamma}$ the HB bound (the light green region delimited by a dotted line) applies. If $g=g_{a\gamma\chi}$ the HB bound vanishes and the RG bound (horizontal solid green line) must be considered.  We show also other bounds, valid for both $g_{a\gamma\gamma}$ and $g_{a\gamma\chi}$. The color code is the same as Fig.~\ref{fig:boundratecdm}.}
	\label{fig:boundg}
\end{figure*}
To circumvent stellar cooling bounds while relaxing anisotropy bounds, we will assume the existence of a small hot dark matter component decaying through a dark portal featuring a very light dark photon ($m_\chi\ll m_a$)~\cite{Kaneta:2016wvf,Kalashev:2018bra,Arias:2020tzl,Hook:2021ous}, $a\rightarrow \chi+\gamma$
 \begin{align}\label{eq:darklag}
     \mathcal{L}\supset \frac{1}{2}g_{a\gamma \chi} a F_{\mu\nu}\tilde{F}^{\mu\nu}_\chi,
 \end{align}
with a decay rate
\begin{equation}
   \Gamma_{a\to\gamma\chi}= \frac{ g_{a\gamma\chi}^2}{32\pi} m_{a}^{3}.
\end{equation}
Notice that the number of photons produced per unit time by the coupling in Eq.~\eqref{eq:darklag} is equal to the ones from the coupling in Eq.~\eqref{eq:lag} for $g_{a\gamma\chi}=g_{a\gamma\gamma} \equiv g$.\footnote{Notice that the numerical factor in the Lagrangian is different from the one used in Ref.~\cite{Kalashev:2018bra}.}
This interaction was assumed in Ref.~\cite{Kalashev:2018bra} as a possible interpretation of an excess observed in the cosmic infrared background by CIBER~\cite{Matsuura:2017lub}, a sounding rocket equipped with infrared cameras. A similar idea has been recently advanced in Ref.~\cite{Nakayama:2022jza} to explain LORRI excess, but no dedicated analysis was carried out.
We constrain the dark portal model using the same strategy discussed in Sec.~\ref{sec:results}, i.e. by requiring that the angular power spectrum $C_l$ in Eq.~\eqref{eq:Cell} must not exceed the upper error bar of any of the data points. Since for this model the number of photons produced per unit time is the same obtained from the coupling in Eq.~\eqref{eq:lag}, the same bounds apply for both models in all the scenarios under consideration. Indeed, in the case of CDM, the production mechanisms are independent of the coupling $g$. Thus, the angular power spectrum $C_l$ is computed using $\rho_a = \rho_{\rm CDM}$ and the CDM power spectrum $P_\delta$ evaluated with the CLASS code, cutting it at the co-moving wave-number $k_T=7\,h$~Mpc$^{-1}$ when the production is dominated by the annihilation of domain walls. In the case of NCDM, the blue axion is thermally produced mainly via pair annihilations in the s channel $e^+ + e^- \rightarrow a + \chi$, while plasmon decays $\gamma^{*} \rightarrow a + \chi$ give a negligible contribution in the early Universe plasma. The freeze-out temperature for the pair annihilation is  \cite{Kalashev:2018bra}
\begin{equation}
T_F \simeq 4.8 \times 10^3~\GeV~\left(\frac{10^{-9}~\GeV^{-1}}{g_{a\gamma\chi}}\right)^2,
\end{equation}
also in this case larger than the EW scale for $g_{a\gamma\chi} \lesssim 10^{-9}~\GeV^{-1}$. Thus, also for the dark portal model we consider NCDM axions with temperature equal to the neutrino one or set by the modified freeze-out scenario described in Sec.~\ref{sec:freezeout}. For both the scenarios, we are agnostic of the production processes and the angular power spectrum $C_l$ is evaluated using $\rho_a$ and $P_{\delta,\,{\rm NCDM}}$ as described in Sec.~\ref{sec:freezeout} (for the freeze-out) and Sec.~\ref{sec:relic} (for the hot relic case), obtaining the same value of $C_l$ for $g_{a\gamma\gamma}=g_{a\gamma\chi}=g$. In Fig.~\ref{fig:boundg} we show the bounds and projected reaches on the coupling $g$, valid for both $g_{a\gamma\chi}$ and $g_{a\gamma\gamma}$. For axions interacting with two photons, the HB bound (horizontal dotted green line) applies, excluding $g_{a\gamma\gamma}\gtrsim 0.65\times 10^{-10}$~GeV$^{-1}$~\cite{Ayala:2014pea} and ruling out the NCDM interpretation of the LORRI excess in all the mass range we consider. On the other hand, this constraint vanishes for the dark portal since Primakoff emission is not possible. In this model, plasmon decays in stars can copiously produce axion-dark photon pairs, affecting the standard stellar evolution. The plasmon decay rate was explicitly derived in Ref.~\cite{Kalashev:2018bra}, and found to be identical to the plasmon decay rate to neutrinos with a magnetic dipole moment $\mu_\nu$, with the substitution $\mu_\nu\rightarrow g_{a\chi\gamma}/2$.\footnote{The difference in the numerical factor comes again from the different definition we have chosen for the dark portal Lagrangian in Eq.~\eqref{eq:darklag}.} The strongest bounds to date on the neutrino magnetic dipole moment comes from the brightness of the tip of the red-giant branch~\cite{Capozzi:2020cbu}. The best galactic target is $\omega$ Centaury, from which $\mu_\nu <1.2 \times 10^{-12}\mu_B$ at 95\% CL with $\mu_B=e/2m_e$ the Bohr magneton, much more stringent than previous bounds, e.g.~\cite{Viaux:2013lha}. Therefore, we find $g_{a\chi\gamma}<7.1\times 10^{-10}\rm GeV^{-1}$ (the horizontal solid green line).  This is the strongest bound on the dark portal in most of the parameter space, and supersedes all previous astrophysical constraints for any mass smaller than $\mathcal{O}(10\,\rm keV)$~\cite{Arias:2020tzl,Hook:2021ous}.

Nevertheless, the NCDM dark portal interpretation of the excess is excluded.
Future anisotropy measurements at 438 nm and 336 nm will represent the strongest probe on the dark portal model for $8~\eV\lesssim m_{a}\lesssim 24~\eV$ in the case of CDM, $8~\eV\lesssim m_{a} \lesssim 20~\eV$ for the freeze-out scenario and $8~\eV\lesssim m_{a} \lesssim 18~\eV$ for the hot relic case.

\section{Line intensity mapping projected reach}
\label{sec:LIM}

We here briefly comment on the promising possibilities of line intensity mapping (LIM)~\cite{Kovetz:2017agg,Bernal:2022jap}. Similarly to the anisotropy measurements previously described, rather than identifying galaxies, LIM aims at measuring the integrated emission of spectral lines from both galaxies and the intergalactic medium (IGM) with small low-aperture instruments. The line-of-sight distribution is then reconstructed through the frequency dependence: the redshift of a targeted line is obtained comparing the observed frequency with the frequency at rest. A decaying relic would potentially show up in future surveys as an ``interloper line''. The approach, based on previous ideas of Refs.~\cite{Grin:2006aw,Gong:2015hke}, was proposed in Ref.~\cite{Creque-Sarbinowski:2018ebl}, and finally applied to realistic forecasts of decaying DM~\cite{Bernal:2020lkd} $a\to \gamma \gamma$ and neutrinos~\cite{Bernal:2021ylz} $\nu_i \to \nu_j\,\gamma$, where $\nu_i$ is an active massive neutrino with mass $m_i$ decaying into a lighter eigenstate $\nu_j$ with mass $m_j$. The possible observables are the LIM power spectrum and the voxel intensity distribution (VID), i.e. the distribution of observed intensities in each voxel, the three-dimensional equivalent of pixels~\cite{Breysse:2016szq,Bernal:2020lkd}. Here we focus on the most powerful projections, which are obtained with VID measurements.

\begin{figure} \includegraphics[width=0.45\textwidth]{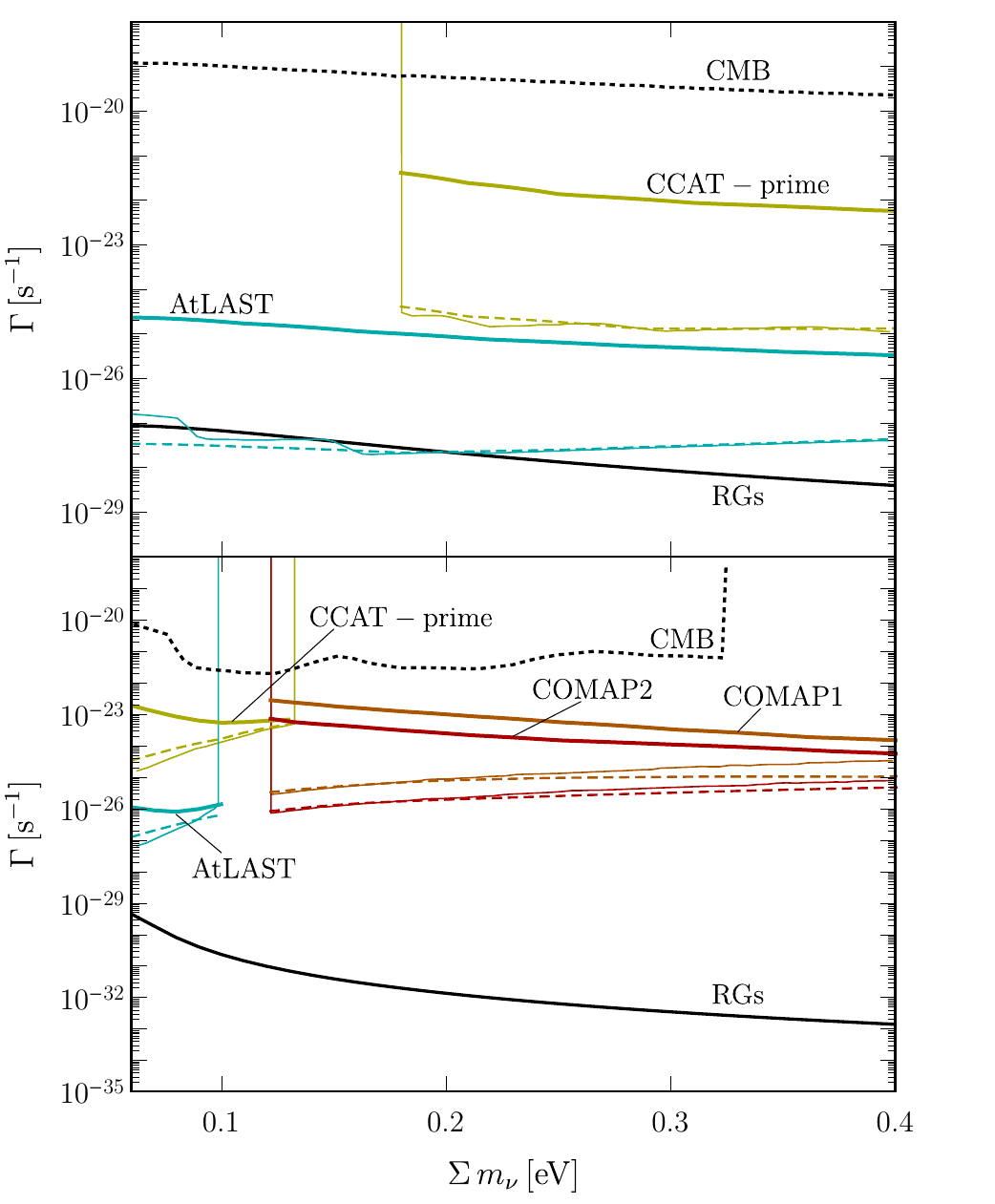}\,
\caption{Projected $95\%$ CL sensitivity on the neutrino lifetime in the case of normal hierarchy as function of the total neutrino mass from VID measurements for all LIM surveys considered in \cite{Bernal:2021ylz} (thin-solid lines), namely COMAP1 (orange),
COMAP2 (red), CCAT-prime (dark yellow) and AtLAST (light blue). The other coloured lines represent the rescaled sensitivities on neutrinos, obtained from the LIM searches for DM decay shown in the published version of Ref.~\cite{Bernal:2020lkd} (dashed) and erratum (thick-solid). We show also CMB~\cite{Aalberts:2018obr}  (dotted black)~limits and the astrophysical bound from RGs~\cite{Capozzi:2020cbu} (solid black). In the upper panel the transition between mass eigenstates $\nu_3 \rightarrow \nu_1$ is considered, while in the lower panel $\nu_2 \rightarrow \nu_1$ is assumed.}
	\label{fig:neutrinos}
\end{figure}

First, in Fig.~\ref{fig:neutrinos} we reproduce the results of Ref.~\cite{Bernal:2021ylz} for the normal hierarchy (the same can be done for the inverted hierarchy). To obtain the dashed curves, we simply rescale the reach of LIM searches for DM decay of Ref.~\cite{Bernal:2020lkd} to the abundance of neutrinos. Namely, we find the projected reach on the neutrino decay rate $\Gamma_\nu$ as a function of the sum of the neutrino masses $\sum m_\nu$, requiring that
\begin{equation}
    \Gamma_\nu n_\nu=\Gamma^{\rm Bernal}_{\rm DM} n_{\rm DM}\,,
\end{equation}
where $n_\nu = 56$~cm$^{-3}$~\cite{ParticleDataGroup:2022pth} is the neutrino number density per flavor, $n_{\rm DM}=\rho_{\rm DM}/m_a$ is the DM number density and $\Gamma^{\rm Bernal}_{\rm DM}$ is the projected reach on DM decay found in Ref.~\cite{Bernal:2020lkd} as a function of the DM mass $m_a$. Here, we rescale the DM mass by replacing $m_a \to (m_i^2 - m_j^2)/m_i$ since the rest-frame energy of the emitted photons is $m_a/2$ in the case of DM decays, and $(m_i^2 - m_j^2)/2m_i$ for neutrino decay. Thus, in the case of a transition $\nu_i \to \nu_j$, the rescaled projected reach on $\Gamma_\nu$ is explicitly given by
\begin{equation}
   \Gamma_\nu\left(\sum m_\nu\right) = \Gamma_{\rm DM}^{\rm Bernal} \left(\frac{m_i^2-m_j^2}{m_i}\right) \frac{\rho_{\rm DM}}{n_{\nu}} \frac{m_i}{m_j^2-m_i^2}\,,
\end{equation}
where $m_j^2 - m_i^2$ is a fixed parameter~\cite{ParticleDataGroup:2022pth} and $m_i$ depends on $\sum m_\nu$.
In the upper panel of Fig.~\ref{fig:neutrinos}, we show the results of our rescaling (the dashed lines) in the case of $\nu_3 \to \nu_1$ decay, with \mbox{$m_3^2 - m_1^2 = 24.53 \times 10^{-4}$~eV$^2$}~\cite{ParticleDataGroup:2022pth}, while in the lower panel we plot the projected reach on neutrino decay for the transition $\nu_2 \to \nu_1$, with \mbox{$m_2^2 - m_1^2 = 0.753 \times 10^{-4}$~eV$^2$}~\cite{ParticleDataGroup:2022pth}. In both cases, we find good agreement between our dashed curves and the results of the dedicated analysis of Ref.~\cite{Bernal:2021ylz} (the thin-solid lines).
This shows that, as for the previously discussed COB anisotropies, the shape of the power spectrum modifies the prediction only negligibly, and one just needs to account for the abundance. Notice, however, that the projected reach for DM decay has been recently computed again by the same authors in an erratum~\cite{Bernal:2020lkd}. Therefore, we revisit the neutrino decay rate reach (see the thick-solid lines) by rescaling the DM reach of the \textit{erratum}.
The reach is less competitive than initially claimed in Ref.~\cite{Bernal:2021ylz}, and weakened to the level of future CMB spectral distortion probes~\cite{Aalberts:2018obr}. Moreover, it is several orders of magnitude larger than astrophysical probes~\cite{Capozzi:2020cbu}, given by the sum of all the possible diagonal and non-diagonal electric and magnetic moments, so the latter are to be interpreted as conservative bounds on the lifetime of a specific neutrino mass eigenstate. On the other hand, the NCDM component can be much more abundant than neutrinos, offering an interesting target to forthcoming LIM searches.

 \begin{figure*}
\includegraphics[width=0.45\textwidth]{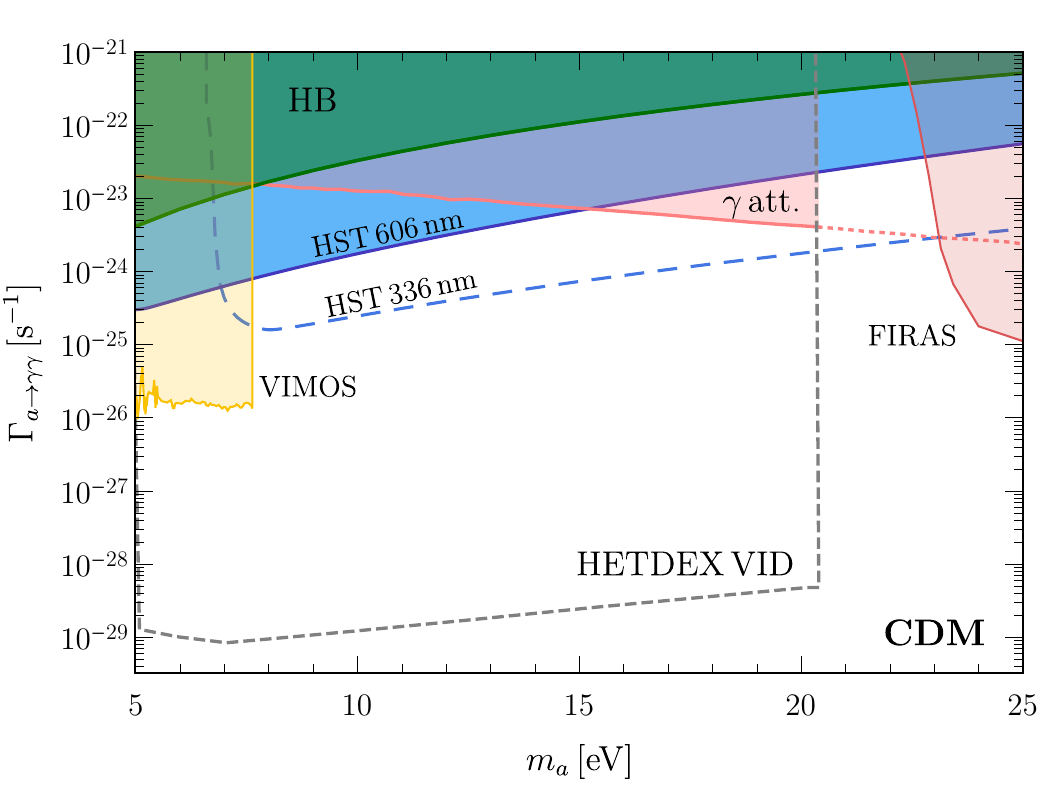}\,
\includegraphics[width=0.45\textwidth]{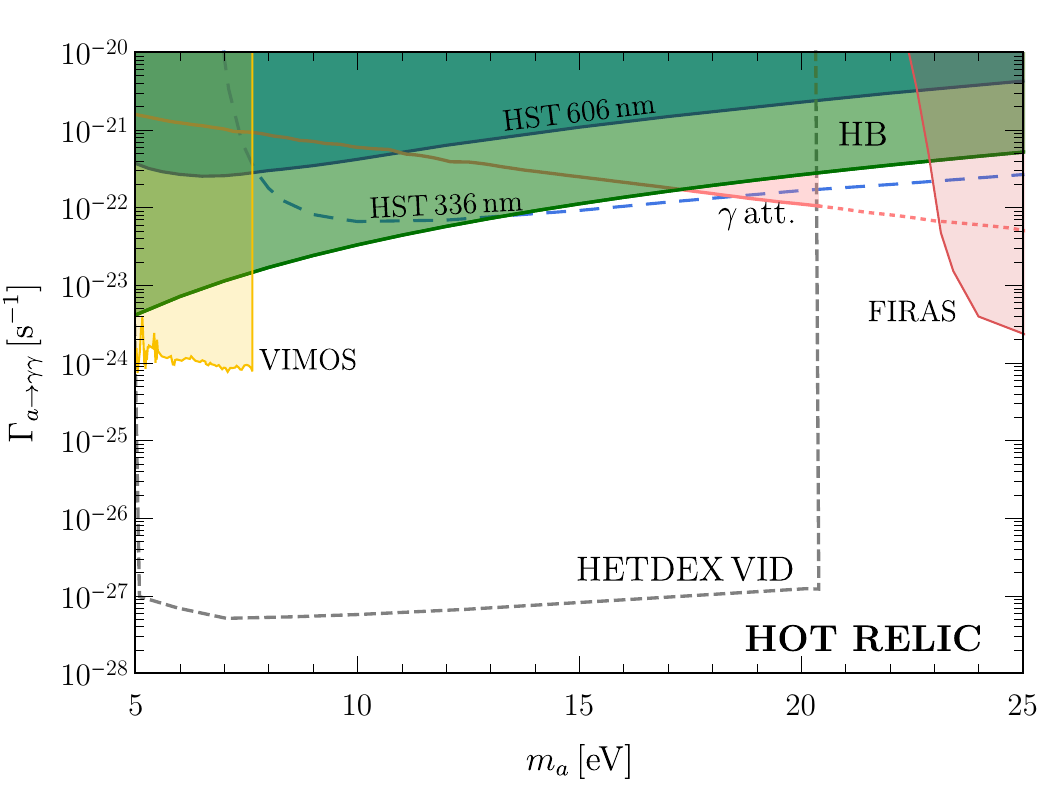}\,
	\caption{Projections on the lifetime of the blue axion for LIM VID (dashed gray line) for CDM (left panel) and the hot relic case (right panel). We show also other bounds and reaches with the same color code of Fig.~\ref{fig:boundratecdm}.}
	\label{fig:VIDrate}
\end{figure*} 
Without further ado, we can obtain the reach for NCDM by simply rescaling the projected reach for CDM by a factor $\Omega_{\rm CDM}/\Omega_a$, with $\Omega_a$ depending on the cosmological scenario. Thus, we compute the LIM projections and show them in the $m_a\,-\,\Gamma$ plane for CDM (taken from the erratum) in the left panel of Fig.~\ref{fig:VIDrate} and the most conservative hot relic case (right panel), together with other bounds (solid lines) and the strongest reaches (dashed lines) on the blue axion. From the updated Ref.~\cite{Bernal:2020lkd} we see that the Hobby-Eberly Telescope Dark
Energy Experiment (HETDEX)~\cite{Hill:2008mv} can probe CDM blue axions with lifetime $10^{29}$~s, and by a simple rescaling argument we find HETDEX can probe also hot relic blue axions with lifetime $10^{27}\,\rm s$.

\section{Discussion and conclusions}
\label{sec:conclusions}
In this paper, we have revisited the parameter space of a radiative decaying Big Bang relic with a mass $m_a\simeq 5-25 \,\rm eV$ that we have dubbed ``blue axion''. We explored the case of the decay to two photons $a\rightarrow \gamma+\gamma$ and the case of decay to a photon and a dark vector, $a\rightarrow \chi+\gamma$.

Existing bounds in this part of the parameter space depend on the production mechanism that dictates both the abundance and the power spectrum, except for stellar bounds (evolution of HB stars for the two-photon coupling, and of RGs for the dark portal). Therefore, we have considered four different scenarios---namely, cold dark matter produced through the misalignment mechanism (with additional contributions from strings and rapidly decaying domain walls), a late-time produced cold dark matter from the annihilation of a long-lived string-wall network, a thermally produced population with temperature set by a freeze-out mechanism, and a hot relic with the temperature of neutrinos and the maximum abundance allowed by structure formation bounds.

We found that cosmic optical background anisotropies are a powerful probe, as the results we obtain depend mostly on the abundance, and only marginally on the power spectrum.
We have proposed the observation of the COB anisotropies with HST at the pivot wavelengths of 438~nm and 336~nm, which has the most competitive discovery reach to date. In the near future, line intensity mapping will be a powerful probe of the blue axion parameter space, probing lifetimes as large as $10^{29}\,\rm s$ or $10^{27}\,\rm s$ assuming the blue axion to be the totality of cold dark matter or a hot relic with the abundance limited by structure formation bounds, respectively. As a byproduct, our analysis also excludes the dark portal model interpretation of the excess recently detected by LORRI on board of the New Horizons mission, lending further credibility to an astrophysical interpretation of the data. However, the experimental directions proposed in our paper will explore presently unprobed coupling strengths. Therefore, blue axions could still be discovered in the nearby future.

\section*{Acknowledgements}
We are grateful to Luca Di Luzio, Maurizio Giannotti and Alessandro Mirizzi for conversations at an early stage of this project. We thank Asantha Cooray for discussions about the the UVIS channels of HST, and Andrea Caputo for a discussion about LIM bounds. We also thank Andrea Caputo, Damiano Fiorillo, Eric Kuflik, Alexander Kusenko, Ciaran O'Hare and Nicholas Rodd for comments on a first draft of this paper. This article is based upon work from COST Action COSMIC WISPers CA21106, supported by COST (European Cooperation in Science and Technology).
The work of PC is supported by the European Research Council under Grant No.~742104 and by the Swedish Research Council (VR) under grants  2018-03641 and 2019-02337. The work of GL is partially supported by the Italian Istituto
Nazionale di Fisica Nucleare (INFN) through the ``Theoretical Astroparticle Physics'' project and by the research
grant number 2017W4HA7S ``NAT-NET: Neutrino and
Astroparticle Theory Network'' under the program PRIN
2017 funded by the Italian Ministero dell’Universit\`a e
della Ricerca (MUR).  EV acknowledges support by the European
Research Council (ERC) under the European Union’s Horizon Europe research and innovation programme (grant agreement No. 101040019). Views and opinions expressed are however those of the author(s) only and do not necessarily reflect those of the European Union.

   \appendix

      \section{Existing constraints on the blue axion}
      \label{app:bounds}
 In this Appendix we comment on all the existing bounds on the blue axion, i.e. from optical telescopes (VIMOS), CMB spectral distortions (FIRAS), the attenuation of the blazar $\gamma$-ray spectra, and stellar cooling (the R parameter, related to the evolution of HB stars).

   \subsection{VIMOS}\label{sec:VIMOS}
   The experiment VIMOS (Visible Multi-Object
Spectrograph), was used to look for optical line emission  in the galaxy clusters Abell $2667$ and $2390$. This observation was used to set constraints on radiative axion decays~\cite{Grin:2006aw}.

Axions thermally produced in the early Universe contribute to the hot DM with the amount shown in Eq.~\eqref{eq:omega}. Today, axions with mass in this range are non-relativistic and bound to galaxy clusters. In the parameter range of interest it is possible that a fraction of the matter in a galaxy cluster is composed by axions.

Axions decay to photons with energy $m_{a}/2$ which is subsequently redshifted until they reach the detector. If the axions
have a cosmological density given $\rho_{\rm NCDM}$ and they compose a significant fraction of the cluster, the intensity from axion decay is
\begin{equation}
I=\omega\frac{\rho_{\rm NCDM}V_{\rm cl}}{m_{a}(1+z_{cl})^{4}4\pi d_{cl}^{2}}\Gamma_{a\to\gamma\gamma},
\label{eq:b1}
\end{equation}
where $V_{\rm cl}$ is the volume of the cluster, and $d_{\rm cl}$ is the distance corresponding to a redshift $z_{\rm cl}$ ($z_{\rm cl}=0.233$ for A 2667 and $z_{\rm cl}=0.228$ for A 2390).

Two galaxy clusters were observed by VIMOS in the period June 27-30, 2003~\cite{Covone:2005uv}.  The signature of decaying axions is an emission line tracing the DM density in the cluster.  The absence of such a signal results in a constraint on axion CDM in the range $4.5~\eV\lesssim m_{a}\lesssim 7.7~\eV$ and $g_{a\gamma\gamma}\lesssim 3\times10^{-12}~{\rm GeV}^{-1}$, as shown in Fig.~\ref{fig:boundg}.

Note that Eq.~\eqref{eq:b1} scales like $\sim g_{a\gamma\gamma}^{2}\Omega_{a}$ and the bounds in Ref.~\cite{Grin:2006aw} (see Fig.~7 therein) were obtained assuming a NCDM with the relic density ${\Omega_{\rm NCDM}\,h^2= m_a/130{\,\rm eV}}$, which is excluded by structure formation bounds~\cite{Diamanti:2017xfo}. Moreover, besides the axion-photon coupling, additional interactions are needed to guarantee this relic density, obtained from Eq.~\eqref{eq:omegancdm} by setting $g_{*,s}=10$. To obtain the constraint in the CDM case shown in Fig.~\ref{fig:boundratecdm} we need to multiply the intensity in Eq.~\eqref{eq:b1} by a factor $\Omega_{\rm CDM}/\Omega_{\rm NCDM}$.\footnote{We mention that the constraint shown in Fig.~3 of Ref.~\cite{Bernal:2022xyi} is incorrect, as it is obtained by rescaling the original results of Ref.~\cite{Grin:2006aw} by a spurious factor $(\Omega_{\rm CDM}/\Omega_{\rm NCDM})^{2}$. This has been also recently corrected in the repository of Ref.~\cite{AxionLimits}.}

      \subsection{FIRAS}

The decay of a massive particle leads to the injection of photons in the primordial bath producing spectral distortions of the CMB. This effect is more efficient when the energy of the produced photon exceeds the neutral hydrogen excitation energy ($13.6$~eV). 
However, any source of distortion is constrained by the fact that the energy spectrum of the CMB is compatible with a perfect blackbody at $\sim2.7$~K~\cite{ParticleDataGroup:2022pth}. 

In this context, in Ref.~\cite{Bolliet:2020ofj}, data from the experiments \textit{COBE}/FIRAS, EDGES and Planck were used to constrain an exotic energy injection in the primordial plasma.
Axions with  $m_{a}\gtrsim 27$~eV and ${g_{a\gamma\gamma}\gtrsim 3\times10^{-12}~\GeV^{-1}}$  are excluded by this argument.

Since the energy injected is proportional to the abundance of decaying DM, the bounds on the rate shown in Fig.~\ref{fig:boundratecdm} can be rescaled to different scenarios by multiplying the squared coupling from Fig.~29 of Ref.~\cite{Bolliet:2020ofj} (valid for CDM) by a factor ${\Omega_{a}/\Omega_{\rm CDM}}$, with $\Omega_{a}$ depending on the considered cosmological scenario.

\subsection{$\gamma$-ray attenuation}
 \begin{figure}
	\includegraphics[width=0.45\textwidth]{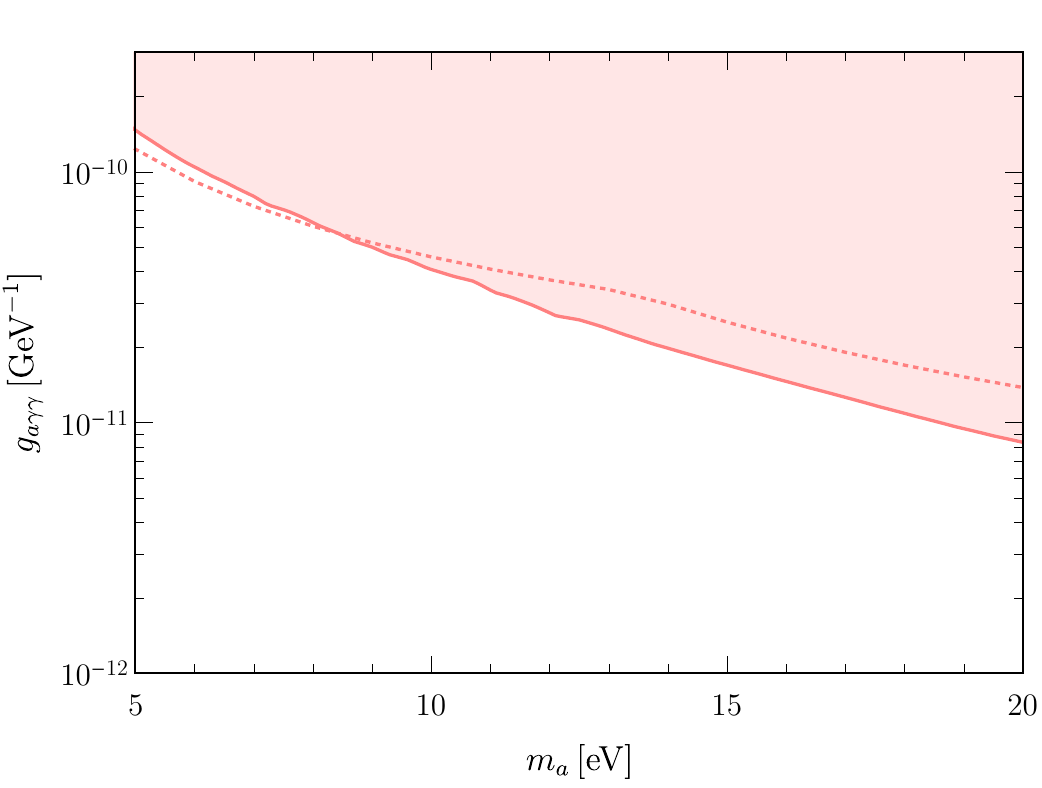}\,
	\caption{The most conservative $\gamma$-ray bound obtained in Ref.~\cite{Bernal:2022xyi} (solid pink line) and the one estimated in this Appendix (dotted pink line).}
	\label{fig:gammaraybound}
\end{figure}  
High-energy $\gamma$ rays from far sources are attenuated by scattering on low-energy photons of the EBL and subsequent pair production, $\gamma+\gamma\rightarrow e^+ + e^-$. One can determine this attenuation as a function of the source redshift and the observed $\gamma$-ray energy through joint analyses of the observed blazar
spectra. In this context, Ref.~\cite{Kalashev:2018bra} advanced the idea that the blazar $\gamma$-ray spectrum is sensitive to the decay of eV-scale axions decaying into photons (see also Ref.~\cite{Korochkin:2019pzr,Korochkin:2019qpe}). In the recent Ref.~\cite{Bernal:2022xyi}, blue axions are constrained by performing a likelihood analysis, comparing the predicted optical depths due to axion decays with optical depth measurements obtained from the observations of almost 800 blazars~\cite{TheFermi-LAT:2016zue,Desai:2019pfl}. More precisely, the blue axion decay contribution is compared with the residual optical depths obtained by subtracting from EBL three standard components, namely the
emission from galaxies at $z<6$, the emission from galaxies
at $z>6$, and the intra-halo light (IHL)—emission from
a faint population of stars tidally removed from galaxies~\cite{Cooray:2012xj,Zemcov:2014eca,Matsumoto:2019stv}. Since the contribution from
galaxies at $z<6$ vastly dominates the EBL, in Ref.~\cite{Bernal:2022xyi} the most conservative bound, represented by the solid line in Fig.~\ref{fig:gammaraybound}, is obtained increasing the uncertainties of the astrophysical EBL from galaxies
at $z < 6$ to saturate the measured EBL for all the considered source redshifts and observed $\gamma$-ray energies.

To give some intuition about the nature of these bounds, here we reproduce the $\gamma$-ray constraints within at most a factor of $\mathcal{O}(1)$ through a simpler, less refined approach, requiring that the optical depth due to axion decay does not exceed the upper error bar of any data point for the optical depth at $z_s=2.4$, the largest source redshift shown in Ref.~\cite{Bernal:2022xyi} (see Fig.~2 therein).

The optical depth is defined in terms of the $\gamma$-ray mean free path $l$ as 
\begin{equation}
\tau= \int_0^{z_{s}}d  z\, \frac{l^{-1}}{(1+z)\,H},
\label{eq:tau}
\end{equation}
where $z_{ s}$ is the source redshift. Let us consider a $\gamma$-ray photon observed on Earth with an energy $E_\gamma$. At redshift $z$, it has energy $\epsilon_{\gamma}=(1+z)E_{\gamma}$, and can potentially scatter on a low-energy photon of the EBL with energy $\epsilon$. 
Electron-positron pair production can occur above the EBL photon energy threshold
\begin{equation}
    \epsilon_{\rm min} = \frac{2m_e^2c^4}{\epsilon_\gamma(1-\mu)},
\end{equation}
where $m_e$ is the electron mass and $\mu$ is the cosine of the incidence angle of the two photons. The pair production cross section is~\cite{Gould:1967zzb},

\begin{equation}
\begin{split}
    \sigma_{\gamma\gamma}(\beta) & =\frac{3\sigma_{\rm T}}{16}\left(1-\beta^2\right)\\
    & \times\left[2\beta\left(\beta^2-2\right)+\left(3-\beta^4\right)\ln \left(\frac{1+\beta}{1-\beta}\right)\right],
\end{split}
\end{equation}
where $\sigma_T=6.65 \times10^{-25}\cm^{2}$ is the Thomson cross-section, $\beta^2=1-2m_e^2/(\epsilon\epsilon_\gamma(1-\mu))$ is the electron velocity in the center-of-mass frame. Thus, the inverse mean free path is given by
\begin{equation}
    l^{-1} = \int_{0}^{\infty}d  \epsilon\left(\frac{d  n}{d  \epsilon}\right)^{\rm dec} \int_{-1}^{1}d  \mu \frac{(1-\mu)}{2}\sigma_{\gamma\gamma}\Theta(\epsilon-\epsilon_{\rm min}),
    \label{eq:mfp}
\end{equation}
where $\Theta$ is the Heaviside function and $(dn/d\epsilon)^{\rm dec}$ is the axion contribution to the EBL, evaluated as

\begin{equation}
     \left(\frac{d  n}{d  \epsilon}\right)^{\rm dec}  = \frac{2\Omega_{a}\rho_{\rm c}\Gamma_a(1+z)^3}{m_a \epsilon H(z_*)}\Theta(z_\star-z).
\label{eq:dnde_dm}
\end{equation}

Here, $\Omega_a$ is the axion density parameter and $z_*\equiv m_a c^2(1+z)/(2\epsilon)-1$ is the redshift of decay that contributes to the photon energy and redshift of interest. For fixed values of the axion mass and coupling, plugging Eq.~\eqref{eq:mfp} into Eq.~\eqref{eq:tau} we obtain the optical depth $\tau$ as a function of the observed $\gamma$-ray photon energy. If one assumes that axions constitute all of the CDM, $\Omega_a = \Omega_{\rm CDM}$, requiring that the computed $\tau$ must not exceed the upper error bar of any of the measured optical depths at $z_s=2.4$~\cite{Bernal:2022xyi}, the dotted line in Fig.~\ref{fig:gammaraybound} is obtained, reproducing the $\gamma$-ray bound in Ref.~\cite{Bernal:2022xyi} within a factor 2 for all masses in the range of interest.

Since the optical depth $\tau$ scales as $\Omega_a\,\Gamma_a$, the bounds on the rate shown in Fig.~\ref{fig:boundratecdm} are obtained by multiplying the CDM constraint (taken from Ref.~\cite{Bernal:2022xyi}) by a factor $\Omega_{\rm a}/\Omega_{\rm CDM}$, with $\Omega_a$ depending on the considered cosmological scenario. Notice that these bounds are completely independent of the relic power spectrum. We also stress that neutral hydrogen absorption (for axion masses above 20.4 eV) can modify the constraints. Hence, we show this region as dotted in all the relevant plots of the main text.

         \subsection{Horizontal branch stars}
        
The axion-photon interaction turns on the production of light axions in stars through the Primakoff process~\cite{Raffelt:1985nk}, altering the stellar evolution. The best astrophysical probes of the axion-photon coupling are Horizontal Branch (HB) stars~\cite{Raffelt:1987yu,Raffelt:1996wa,Raffelt:2006cw}. Axions produced by Primakoff processes reduce the lifetime of HB stars, without affecting the previous red giant (RG) phase because of the large electron degeneracy and the high plasma frequency which prevents an efficient axion production in degenerate stars~\cite{Raffelt:1987yu}.

Therefore, the axion-photon interaction can be constrained by means of the $R$ parameter, $R= {N_{\rm HB}}/{N_{\rm RGB}}$, which compares the number of stars in the HB ($N_{\rm HB}$) and RGB ($N_{\rm RGB}$) phases, or equivalently, the duration of these phases. 
This observable is also sensitive to the helium mass fraction $Y$, introducing a degeneracy between $g_{a\gamma\gamma}$ and $Y$ because a lower lifetime due to the presence of axions could be compensated by the increase of the helium abundance.
Recent observations obtained a value of the $R$ parameter equal to $R_{\rm ave}=1.39\pm 0.03$~\cite{Salaris:2004xd}. By comparing the $R$ parameter computed through stellar simulations including axions with the measured one, the constraint derived on axions is~\cite{Ayala:2014pea,Straniero:2015nvc}
\begin{equation} 
 g_{a\gamma\gamma}<0.65\times 10^{-10}\,\ \textrm{GeV}^{-1} \,\  \,\ (95 \% \,\ \textrm{CL})  \;.
 \label{Eq:2sigma}
\end{equation}
A similar bound can be obtained by requiring the exotic energy $\epsilon_{\rm exotic}$ emitted per
unit time and mass to be $\epsilon_{\rm exotic}\lesssim 10~\rm erg\, g^{-1}\, s^{-1}$, to be computed assuming a one-zone model for the core of HB stars, $T\simeq 8.6 \,\rm keV$ and $\rho\simeq 10^4 \, \rm g\, cm^{-3}$ for the temperature and density respectively (see, e.g.,~\cite{Raffelt:1987yb,Raffelt:1996wa,Cadamuro:2011fd}). 

Recently, a different observable has been proposed, namely the ratio of asymptotic giant branch (AGB) to HB stars (the $R_2$-parameter)~\cite{Dolan:2022kul}. This parameter was measured thanks to the HST photometry of 48 globular clusters, obtaining $R_{2}=0.117\pm0.005$~\cite{Lattanzio2}. This observable is known to be quite insensitive to the initial helium mass fraction, in contrast with the $R$ parameter~\cite{Lattanzio2}. The constraint found in this way improves the previous one to the value $g_{a\gamma\gamma}<0.47\times 10^{-10}\GeV^{-1}$.
Evidences from asteroseismology might help to improve this modeling~\cite{Lattanzio1}, leading to a further improvement of the bound to $g_{a\gamma\gamma}<0.34\times 10^{-10}\GeV^{-1}$~\cite{Dolan:2022kul}. We mention that, despite being more stringent that the one based on the $R$ parameter, the bound from the $R_{2}$ parameter is affected by larger uncertainties related to the description of convective phenomena.

 \section{Details about the Hubble Space Telescope} 
   \label{app:HST}
 
 \begin{figure}
	\includegraphics[width=0.45\textwidth]{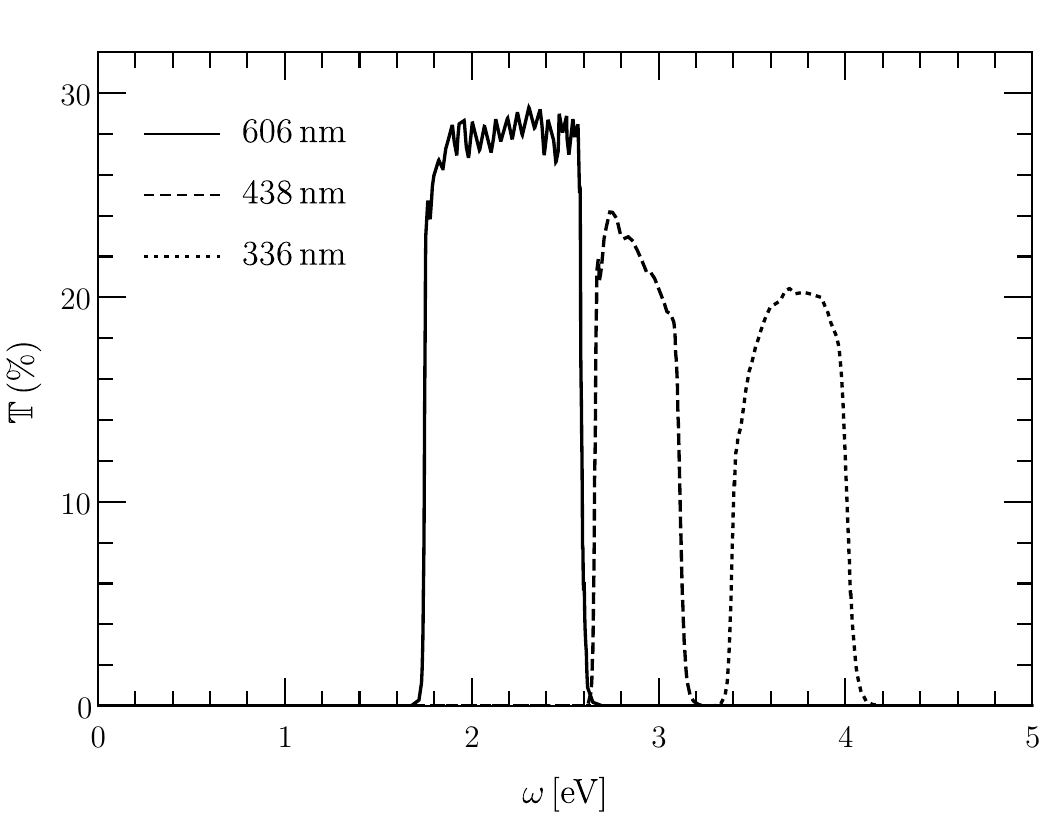}\,
	\caption{Throughput UVIS2 for the filters F606W (solid), F438W (dashed) and F336W (dotted), taken from Ref.~\cite{WFC3}.}
	\label{fig:throughput}
\end{figure}

The Hubble Space Telescope (HST) is a space telescope, in orbit since 1990. It is one of the semiformal group of NASA's ``Great Observatories'' together with the Compton Gamma Ray Observatory, the Chandra X-Ray Observatory, and the Spitzer Space Telescope. HST observes the electromagnetic spectrum in the ultraviolet (UV), visible, and near-infrared (NIR) wavelengths, through four currently active instruments, namely the Advanced Camera for Surveys (ACS), the Cosmic Origins Spectrograph (COS), the Space Telescope Imaging Spectrograph (STIS) and the Wide Field Camera 3 (WFC3). The WFC3, combining two ultraviolet/visible (UVIS) CCDs with a NIR HgCdTe array, is capable of direct, high-resolution imaging over the entire wavelength range 200 to 1700 nm, with the UVIS channel sensitive to $200~{\rm nm} \lesssim \lambda \lesssim 1000~{\rm nm}$, and the IR channel $800~{\rm nm} \lesssim \lambda \lesssim 1700~{\rm nm}$. In this work, we are interested in observations through the UVIS channel, in which the detectors are two $4096 \times 2051$ pixel CCDs (namely UVIS 1 and UVIS 2), butted together to yield a $4096  \times 4102$ light-sensitive array with a $\sim 31$ pixel (1.2 arcsec) gap. 
  In order to characterize the anisotropy spectrum as seen in the detector, we consider three wide-band filters from UVIS 2,\footnote{Even though we use as a benchmark UVIS 2, the results of this work would not sensitively change using UVIS 1, since the two detectors are characterized by almost equal parameters~\cite{WFC3}.} namely 606 nm, 438 nm, and 336 nm, with the pivot-frequency $\omega_{ \rm pivot}$ in Table~\ref{tab:filters} and the throughput $\mathbb{T}(\omega)$ shown in Fig.~\ref{fig:throughput}, defined as the number of detected counts/s/$\rm cm^2$ of telescope area relative to the incident flux in photons/s/$\rm cm^2$. 
  In this way, the normalized throughput function used in Eq.~\eqref{eq:Cell} is
\begin{align}
    \epsilon(\omega)=\frac{\mathbb{T}(\omega)}{\int_0^\infty \frac{d\omega}{\omega} \mathbb{T}(\omega)},
\end{align}
where the integral in the denominator is the UVIS 2 efficiency as defined in Ref.~\cite{WFC3} and reported in Table~\ref{tab:filters}.

   \begin{table}[b!]
    \centering
    \begin{tabular}{|c|c|c|}
    \hline
   Filter & $\omega_{\rm pivot}$ (eV) & Efficiency \\
    \hline
F336W & 3.70 & 0.0313 \\
F438W & 2.87 & 0.0347\\
F606W & 2.11 & 0.1093 \\
    \hline
    \end{tabular}
\caption{Pivot frequency $\omega_{\rm pivot}$ and efficiency for the three wide-band filters F336W, F438W and F606W, taken from Ref.~\cite{WFC3}.}
\label{tab:filters}
\end{table} 

 \vspace*{-0.1cm}
\bibliographystyle{bibi}
\bibliography{biblio.bib}

\newpage
\clearpage

\onecolumngrid
\section*{Note added}

 \begin{figure*}
 \includegraphics[width=0.49\textwidth]{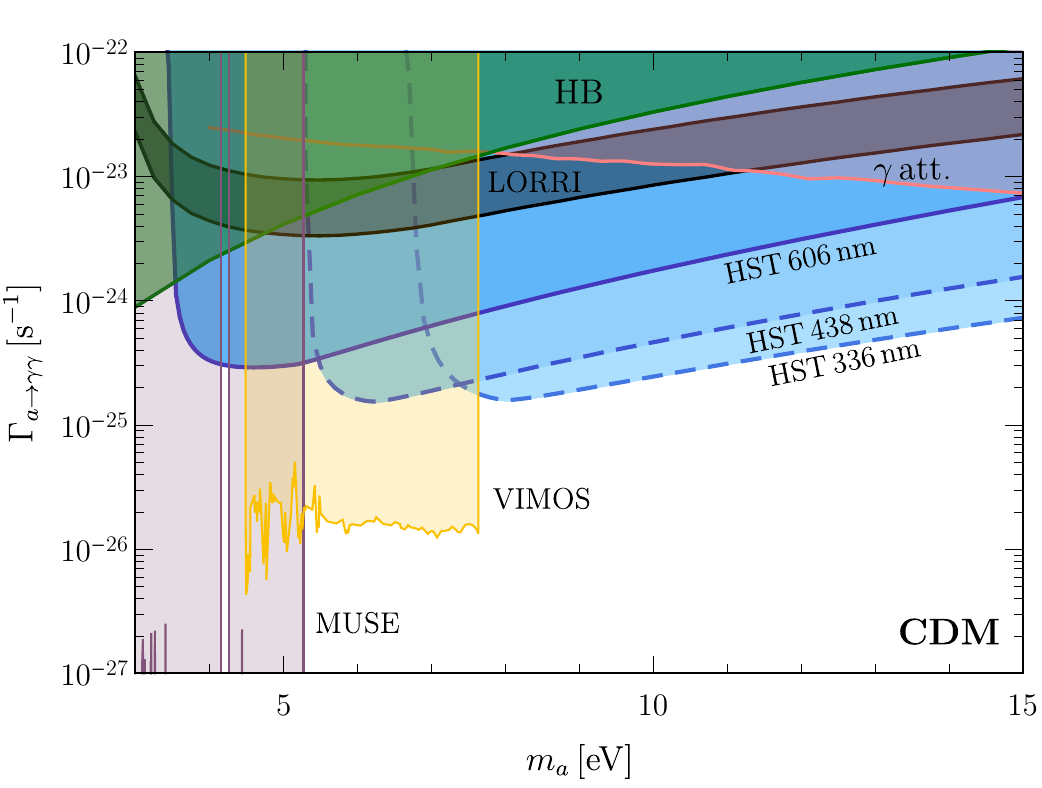}\,
\includegraphics[width=0.49\textwidth]{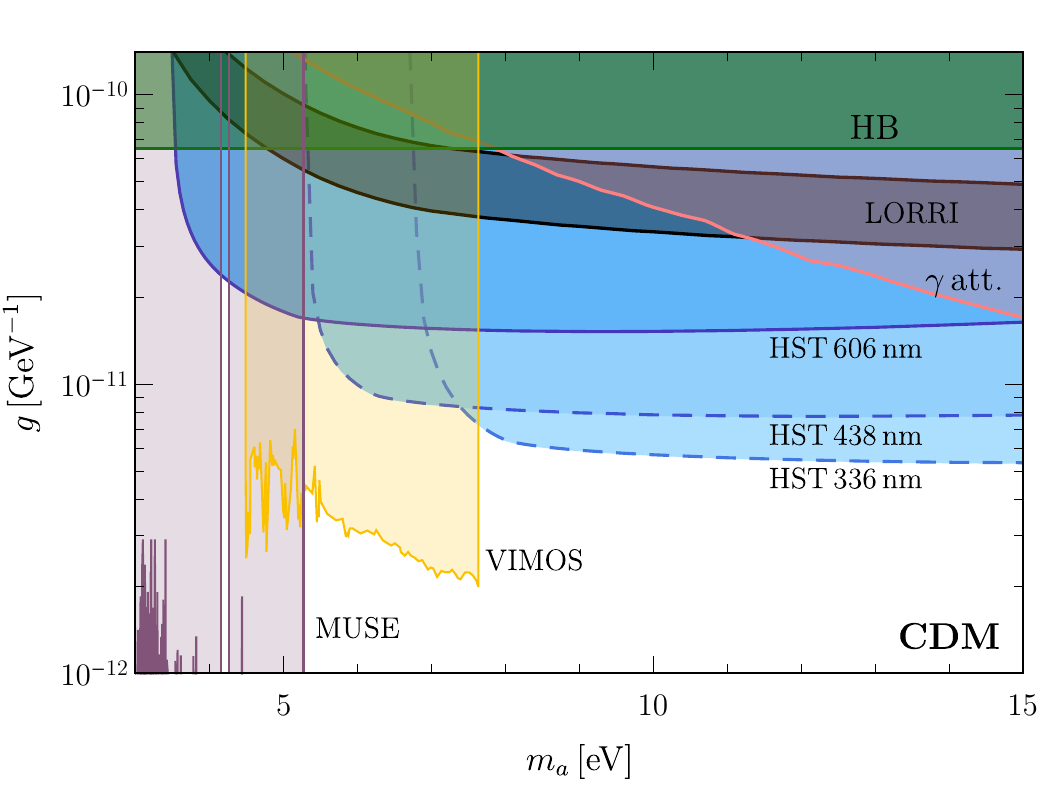}\,
\caption{Bounds and projected reaches on the lifetime (left panel) and coupling $g$ (right panel) of the blue-axion cold dark matter. The $606\,\rm nm$ bound (solid blue lines) degrades at $m_a \lesssim 4$~eV. The forecasts of our proposed observations at $438\,\rm nm$ and $336\,\rm nm$ are given with dashed blue lines. Other constraints are also shown: HB stars~\cite{Ayala:2014pea} (green), MUSE~\cite{Todarello:2023hdk} (purple), VIMOS~\cite{Grin:2006aw} (yellow), FIRAS~\cite{Bolliet:2020ofj} (red), and $\gamma$-ray attenuation~\cite{Bernal:2022xyi} (pink) bounds. The black band identifies the $95\%$ CL excess detected by LORRI~\cite{Lauer:2022fgc,Bernal:2022wsu}.}
	\label{fig:bound_3eV}
\end{figure*} 

In the main text we showed the parameter space for blue axion dark matter in the mass range $m_a \simeq 5 - 25$~eV (Fig.~\ref{fig:boundratecdm}). We have chosen to include a note that addresses smaller masses for the community to reference. The constraint from the Hubble Space Telescope (HST) measurements of Cosmic Optical Background (COB) anisotropies at 606 nm is indeed relevant also for $m_a \lesssim 5$~eV. Axion masses $m_a \simeq 2.7-5.3$~eV are constrained by Multi Unit Spectroscopic Explorer (MUSE) observations of five dwarf spheroidal galaxies, namely Leo T, Sculptor, Eridanus 2, Grus 1, and Hydra II~\cite{Todarello:2023hdk}. However, due to a blocking filter used in the experiment, there is a gap in the bounds from MUSE at $m_a = 4.15-4.27$~eV. 
As shown in the left panel of Fig.~\ref{fig:bound_3eV} for $m_a = 4.2$~eV one can exclude $\Gamma_{a\to\gamma\gamma}\gtrsim 3.03\times 10^{-25}\,\rm s^{-1}$, corresponding to $g\gtrsim 2.33\times 10^{-11}\,\rm GeV^{-1}$ (see right panel), implying that HST measurements lead to the currently strongest bound for $m_a = 4.15-4.27\,\rm eV$.

\end{document}